\documentclass[aps,prd,preprint,nofootinbib,superscriptaddress
]{revtex4-2}
\usepackage{amsfonts}
\usepackage{mathrsfs}
\usepackage{graphicx}
\usepackage{subfigure}
\usepackage{epsfig}
\usepackage{amsmath}
\usepackage{amssymb}
\usepackage{multirow}
\usepackage{booktabs}
\usepackage{array}
\usepackage{tabularx}
\usepackage{slashed}
\usepackage{ulem}
\usepackage{color}

\DeclareGraphicsExtensions{.eps}
\begin{document}
\baselineskip 20pt
%%%%%%%%%%%%%%%%%%%%%%%%%%%%%%%%%%%%%%%%%%%%%%%%%%%%%%%%%%%%%%%%%%%%%%
\title{Production of leptonium in heavy quarkonium decays}

\author{Yu-Han Zhao}
\email{zyh329@mail.sdu.edu.cn}
\affiliation{School of Physics, Shandong University, Jinan, Shandong 250100, China}

\author{Jun Jiang}
\email{jiangjun87@sdu.edu.cn, corresponding author}
\affiliation{School of Physics, Shandong University, Jinan, Shandong 250100, China}

\date{\today}

\begin{abstract}
Lepton pairs with opposite charges can form bound states known as ``leptonium'' through quantum electrodynamic interactions. Heavy quarkonia such as $J/\psi$ are abundantly produced at facilities like BESIII and the future Super Tau-Charm Facility (STCF).  
In this work, we investigate leptonium production in heavy quarkonium decays, specifically focusing on the processes ${\cal{Q}} \longrightarrow (l_1^+ l_2^-)[n] +\gamma$ ($l_{1,2}= \tau,\, \mu,\, e$) and ${\cal{Q}} \longrightarrow (l_1^+ l_2^-)[n] + l_1^- l_2^+$. Here, ${\cal{Q}}$ denotes the heavy quarkonium $J/\psi$ or $\Upsilon$, while $n=$ ${^1S_0}$ or $^3S_1$ corresponds to para-leptonium and ortho-leptonium, respectively.  
With an annual production of $3.4 \times 10^{12}$ $J/\psi$ events at STCF, there is significant potential to observe positronium $(e^+e^-)$, muonium $(\mu^+e^-)$, and dimuonium $(\mu^+\mu^-)$. In particular, the ortho-dimuonium $(\mu^+\mu^-)[^3S_1]$ may be discovered at the future STCF, with an inclusive branching fraction of $Br(J/\psi \longrightarrow (\mu^+\mu^-)[^3S_1] +X) = 1.5 \times 10^{-12}$.  

\vspace {2mm} 
%\noindent {PACS number(s): 12.38.Bx, 12.39.Jh, 14.40.Pq, 14.70.Bh}
\end{abstract}

\maketitle

\newpage
%%%%%%%%%%%%%%%%%%%%%%%%%%%%%%%%%%%%%%%%%%%%%%%%%%%%%%%%%%%%%%%%%%%%
\section{INTRODUCTION}
\label{sec:introduction}
%%%%%%%%%%%%%%%%%%%%%%%%%%%%%%%%%%%%%%%%%%%%%%%%%%%%%%%%%%%%%%%%%%%%
%introduction of Leptonium
Leptons with opposite charges can form bound states under quantum electrodynamics (QED) interactions, known as leptonium. Among the six possible leptonia $(e^+e^-),\,(\mu^\pm e^\mp),\,(\mu^+ \mu^-),\,(\tau^\pm e^\mp),\,(\tau^\pm \mu^\mp),\,(\tau^+ \tau^-)$, only the first two—positronium \cite{Deutsch:1951zza} and muonium \cite{Hughes:1960zz}—have been observed. In addition to these two, the exotic $(\pi \mu)$ atom \cite{Coombes:1976hi} and dipositronium molecule $(e^+e^-)(e^+e^-)$ \cite{Cassidy:2007blx} have also been successfully produced.  
The spectroscopic notation $n\, ^{2S+1} L_J$ is used to describe the quantum numbers of leptonium. Here, $n$ denotes the principal quantum number, $S$ is the spin, $L$ represents the orbital angular momentum (with $L=0,1,2,\dots$ corresponding to S, P, D, $\dots$ waves), and $J=L+S$ is the total angular momentum. Specifically, $J=0$ and $J=1$ correspond to para-leptonium and ortho-leptonium states, respectively. The dominant decay channels for para-leptonium and ortho-leptonium are diphotons and dileptons, respectively, and these have been calculated up to next-to-next-to-leading order \cite{dEnterria:2022alo}.  
In this paper, we focus on the ground states: the spin-singlet $1\,^1S_0$ and spin-triplet $1\,^3S_1$, whose $J^{PC}$ quantum numbers are $0^{-+}$ and $1^{--}$, respectively. Note that dimuonium $(\mu^+\mu^-)$ and ditauonium $(\tau^+\tau^-)$ are also referred to as ``true muonium'' and ``true tauonium'' in the literature.

%introduction of heavy quakronium
Heavy quarkonia are bound states of heavy quark-antiquark pairs $Q\bar{Q}$ (where $Q$ denotes $c$ or $b$ quarks) formed under quantum chromodynamics (QCD) interactions. To date, the $J^{PC}=1^{--}$ charmonium $J/\psi$ and bottomonium $\Upsilon$ can be produced in large quantities. The BESIII experiment has accumulated $10^{10}$ $J/\psi$ events \cite{BESIII:2021cxx}, and the future Super Tau-Charm Facility (STCF) is expected to achieve an annual production of $3.4 \times 10^{12}$ $J/\psi$ events \cite{Achasov:2023gey}. Additionally, the Belle experiment has accumulated $1.02\times 10^8$ $\Upsilon$ events \cite{Belle:2012iwr}. These extensive datasets offer excellent opportunities to explore leptonium production in quarkonium decays. In this paper, we investigate leptonium production in the decays of $J/\psi$ and $\Upsilon$.

%Motivation
The motivation for studying leptonium production in heavy quarkonium decays is threefold. First, the large $J/\psi$ datasets at BESIII and the future STCF offer a promising opportunity to observe dimuonium $(\mu^+\mu^-)$ for the first time. Second, the pure leptonic decays of heavy quarkonia feature clean backgrounds, making them ideal for investigating lepton flavor universality (LFU). Third, anomalies in positronium $(e^+e^-)$ production may provide clues about the proposed $X(17)$ particle, which was ``discovered'' in the electron-positron angular correlation of the isoscalar transition of $^8Be$ \cite{Krasznahorkay:2015iga}.

%mechanism of leptonium production
Broadly speaking, there are three primary mechanisms for the production of leptonium, which have been rigorously studied in searches for dimuonium and ditauonium.  
First, leptonium can be produced through several processes in hadronic or nuclear collision environments: photoproduction via $\gamma A \rightarrow (l^+l^-)A$ in $pp$, $pA$, and $AA$ collisions (where $A$ denotes nuclei or atoms with charge $Ze$) \cite{Gevorkian:1998xj,Yu:2013uka,Francener:2021wzx}; two-photon fusion via $\gamma\gamma \rightarrow (l^+l^-)$ in ultraperipheral $AA$ collisions \cite{Francener:2021wzx,Ginzburg:1998df,Azevedo:2019hqp,Yu:2022hdt,dEnterria:2022ysg,dEnterria:2023yao}; electroproduction via $eA \rightarrow eA (l^+l^-)$ \cite{Gevorkian:1998xj}; bremsstrahlung production in $eA$ collisions or fixed-target experiments \cite{Gevorkian:1998xj}; and quark-antiquark annihilation via $q\bar{q} \rightarrow (l^+l^-) g$ in quark-gluon plasma \cite{Chen:2012ci}.  
Second, leptonium can also be produced in $e^+e^-$ colliders via two-photon fusion or $s/t$-channel processes \cite{dEnterria:2022ysg,dEnterria:2023yao,Moffat:1975uw,Brodsky:2009gx,Gargiulo:2023tci}. Leptonium produced via two-photon fusion must be in the $^1S_0$ state due to the Landau-Yang selection rule \cite{Landau:1948kw,Yang:1950rg}, while that produced through photon-mediated $s$-channels must be in the $^3S_1$ state due to $C$-parity conservation.  
Third, unlike the first two direct production mechanisms, leptonium can be searched for indirectly in the rare decays of mesons, which can be abundantly produced at current and future colliders. Examples include $\eta \rightarrow (\mu^+\mu^-) \gamma$ \cite{CidVidal:2019qub}, $K_L \rightarrow (\mu^+\mu^-) \gamma$ \cite{Ji:2017lyh}, $B \rightarrow K^{(*)} + (l^+l^-)$ \cite{Fael:2018ktm}, $J/\psi \rightarrow (\mu^+\mu^-) \gamma$ \cite{Dai:2024yhy}, and $H \rightarrow \gamma/Z + (l^+l^-)$ \cite{dEnterria:2023wjq,Martynenko:2024rfj}.  

%theoretical study on leptonium production
In recent years, the production of dileptonium $(l^+l^-)$ (where $l=e,\mu,\tau$) has been extensively studied. Positronium $(e^+e^-)$ has been thoroughly investigated for precise quantum electrodynamics (QED) tests \cite{Karshenboim:2005iy} and for searches for violations of space-time CPT symmetries \cite{Bernreuther:1988tt,Yamazaki:2009hp}. Following the discovery of the process $e^+e^- \longrightarrow \mu^+\mu^-$, the reaction $e^+e^- \longrightarrow (\mu^+\mu^-)$ has been considered for dimuonium studies \cite{Moffat:1975uw}.  
In Ref. \cite{Brodsky:2009gx}, the authors propose two methods to detect dimuonium at electron-positron colliders: $e^+e^- \longrightarrow (\mu^+\mu^-)$ for para-dimuonium and $e^+e^- \longrightarrow (\mu^+\mu^-)+\gamma$ for both para-dimuonium and ortho-dimuonium. This proposal was further developed in Ref. \cite{Bogomyagkov:2017uul}. The potential for discovering $^3S_1$ true muonium at LHCb has been explored through the process $\eta \rightarrow \gamma (\mu^+ \mu^-)$ with subsequent $(\mu^+ \mu^-) \longrightarrow e^+e^-$ decay in Ref. \cite{CidVidal:2019qub}.  
Very recently, Dai {\it et al.} proposed the production of dimuonium in radiative $J/\psi$ decays, $J/\psi \longrightarrow (\mu^+\mu^-) +\gamma$, and found that dimuonium detection is feasible at the future STCF \cite{Dai:2024yhy}. The electromagnetic production of dimuonium at RHIC and LHC in relativistic heavy ion collisions is discussed in Ref. \cite{Ginzburg:1998df}.  
For ditauonium $(\tau^+ \tau^-)$, its spectroscopy (including energy levels, lifetime, and partial decay widths, {\it etc.}) has been thoroughly studied by d'Enterria {\it et al.} in Ref. \cite{dEnterria:2022alo}. They also meticulously considered the prospects for ditauonium discovery at electron-positron and hadronic colliders \cite{dEnterria:2022ysg,dEnterria:2023yao}, with their results indicating that ditauonium could be observed at FCC-ee, STCF, and LHC.  
In addition, the production of leptonium in $B$ meson decays $B \rightarrow K^{(*)} + (l^+ l^-)$ (where $l=\tau,\, \mu,\,e$) is discussed in Ref. \cite{Fael:2018ktm}. The production of dileptonium in Higgs boson decays $H \rightarrow \gamma+(l^+ l^-)$ and $H \rightarrow Z+(l^+ l^-)$ has also been calculated \cite{dEnterria:2023wjq,Martynenko:2024rfj}.

%This paper
In this paper, we investigate the production of leptonium in heavy quarkonium decays. Non-relativistic QED (NRQED) and non-relativistic QCD (NRQCD) are employed to describe leptonium and heavy quarkonium, respectively.  
For $J/\psi$ decays, we focus on the processes $J/\psi \rightarrow (l^+ l^-)[^1S_0] + \gamma$ and $J/\psi \rightarrow (l_1^+ l_2^-)[n] + l_1^- + l_2^+$ (where $l_{1,2} =\mu,\,e$), with $n$ corresponding to either the $^1S_0$ or $^3S_1$ state. In the case of $\Upsilon$ decays, we explore all six leptonia, as ditauonium and tauonium are kinematically allowed in this context.  
In our analysis, we discuss both the branching fractions and differential distributions of these processes.

%outlines
The rest of the paper is organized as follows. 
In section \ref{sec:charmleptonium}, we study the leptonium production of positronium, dimuonium and muonim in $J/\psi$ decays. 
In section \ref{sec:bottomleptonium}, we explore the leptonium production in $\Upsilon$ decays, especially the production of ditauonium and tauonium. 
Sec.\ref{sec:summary} is reserved for a summary.

%\newpage
%%%%%%%%%%%%%%%%%%%%%%%%%%%%%%%%%%%%%%%%%%%%%%%%%%%%%%%%%%%%%%%%%%%%
\section{Leptonium production in charmonium decays}
\label{sec:charmleptonium}
%%%%%%%%%%%%%%%%%%%%%%%%%%%%%%%%%%%%%%%%%%%%%%%%%%%%%%%%%%%%%%%%%%%%

In this section, we first introduce the frameworks of non-relativistic quantum electrodynamics (NRQED) and non-relativistic quantum chromodynamics (NRQCD). We then investigate several production processes in $J/\psi$ decays: the production of dileptons, the production of dileptons accompanied by a photon, the exclusive production of leptonium plus a photon, and the inclusive production of leptonium. Due to kinematic constraints, only positronium $(e^+e^-)[n]$, muonium $(\mu^\pm e^\mp)[n]$, and dimuonium $(\mu^+\mu^-)[n]$ can be produced in $J/\psi$ decays.

%%%%%%%%%%%%%%%%%%%%%%%%%%%%%%%%%
\subsection{Non-relativistic QED/QCD}
\label{subsec:NRQCD}
%%%%%%%%%%%%%%%%%%%%%%%%%%%%%%%%%

The Bohr radius of the ground state of leptonium is given by $a_0=2/(\alpha m_l)$, where $m_l$ denotes the lepton mass and $\alpha$ is the fine structure constant. The velocity of each lepton in the $n$-th Bohr orbit is $\beta = 1/(n m_l a_0) = \alpha/(2n)$ \cite{dEnterria:2022alo}.  
The low velocity of constituent leptons in the leptonium rest frame justifies the application of non-relativistic bound-state perturbative theory, specifically non-relativistic QED (NRQED) \cite{Caswell:1985ui}. In the threshold region, constituent leptons interact via the exchange of Coulomb photons. The effect of such interactions is of order $\alpha/\beta$, which necessitates resummation to enhance the convergence of perturbative expansions. This is achieved by solving the Green function of the Schr$\ddot{\mathrm{o}}$dinger equation with the Coulomb potential $-\alpha/r$.  
The QED Coulomb scattering of non-relativistic particles was discussed early on by Sommerfeld, Sakharov, and Schwinger \cite{Sommerfeld:1931qaf,Sakharov:1948plh,Schwinger:1973rv}. The explicit form of the imaginary part of the Green function, $\mathrm{Im} G_{E+i\Gamma}(0,0)$, which accounts for bound-state effects involving constituent particles with non-vanishing decay widths, can be found in Refs. \cite{Fadin:1987wz,Fadin:1990wx}. In these works, Fadin and Khoze discussed the threshold behavior for the production of heavy top quark pairs, assuming Coulomb gluon exchanges between top quarks with non-vanishing decay widths.  
An explicit example of applying this method to study dimuonium production in $J/\psi$ decays is provided in Ref. \cite{Dai:2024yhy}.

At leading order (LO), the binding energy (or energy level) $E_n$ and the squared wavefunction $|\Psi_{n}(r=0)|^2$ at the origin for leptonium with principal quantum number $n$ can be determined by solving the non-relativistic Schr$\ddot{\mathrm{o}}$dinger equation with a Coulomb potential:
\begin{align}
    E_n &= -\frac{\alpha^2 m_{l}}{4n^2}, \\
    \Psi_{nS}^2(r=0) &= \frac{(\alpha m_l)^3}{8\pi n^3}.
\end{align}
Here, the subscript $nS$ denotes the $S$-wave ground state with principal quantum number $n$.

In addition to the Green function approach, another method for describing bound states—widely employed in non-relativistic QCD (NRQCD)—combines spin projectors with the wavefunction at the origin. The NRQCD framework was developed by Bodwin, Braaten, and Lepage to study heavy quarkonia, which are bound states of heavy quark-antiquark pairs \cite{Bodwin:1994jh}.  
Similar to leptonium described by NRQED, constituent heavy (anti-)quarks in heavy quarkonia move non-relativistically in the rest frame of the quarkonium, but interact via the exchange of soft gluons carrying QCD color charge. The typical squared velocity $v^2$, used as the small expansion parameter in perturbative calculations, is $\sim 0.3$ for charmonia and $\sim 0.1$ for bottomonia.  
Within the NRQCD factorization formalism, production rates are factorized into short-distance coefficients and long-distance matrix elements. The short-distance coefficients are perturbatively calculable via Feynman diagram computations, while the long-distance matrix elements are non-perturbative yet universal parameters. These matrix elements can be determined through methods such as lattice calculations \cite{Bodwin:1996tg}, potential models \cite{Eichten:1995ch}, and experimental measurements.  
In potential models, long-distance matrix elements are related to the squared Schr$\ddot{\mathrm{o}}$dinger wavefunction $|\Psi_{n}(r=0)|^2$ at the origin, which in turn connects to the radial wavefunction $R_{n}(r=0)$ at the origin. For $S$-wave states, this relationship is given by:
\begin{equation}
    |\Psi_{nS}(r=0)|^2 = |R_{nS}(r=0)|^2/(4\pi).
\end{equation}

In NRQCD, the bound-state effect is incorporated via spin projectors for the two heavy constituent quarks. The spin projectors for the heavy quark pair $Q\bar{Q}^\prime$ are given by Ref. \cite{Petrelli:1997ge} as follows:
\begin{align}
\Pi_{0} & =\frac{\sqrt{M} \Psi_{n}(0)}{4 m_{Q} m_{Q^{\prime}}}\left(p_{\bar{Q}^\prime}-m_{\bar{Q}^{\prime}}\right) \gamma^5\left(p_{Q}+m_{Q}\right) \otimes\frac{\delta_{ij}} {\sqrt{N_c}}, \\
\Pi_{1} & =\epsilon_\alpha\left(p\right) \frac{\sqrt{M} \Psi_{n}(0)}{4 m_{Q} m_{Q^{\prime}}}\left(p_{\bar{Q}^\prime}-m_{\bar{Q}^{\prime}}\right) \gamma^\alpha\left(p_{Q}+m_{Q}\right) \otimes\frac{\delta_{ij}} {\sqrt{N_c}},
\end{align}
corresponding to spin-zero and spin-one states, respectively.  
Here, $p = p_{Q} + p_{\bar{Q}^\prime}$ and $M = m_{Q} + m_{\bar{Q}^{\prime}}$ denote the momentum and mass of the heavy quarkonium, respectively; $p_{\bar{Q}^\prime} = \frac{m_{\bar{Q}^\prime}}{M} p - q$ and $p_{Q} = \frac{m_Q}{M} p + q$ are the momenta of the two constituent quarks; $m_{Q}$ and $m_{\bar{Q}^{\prime}}$ are their respective masses; $q$ is the relative momentum between the constituent quarks in the heavy quarkonium rest frame; $\epsilon_\alpha(p)$ is the polarization vector for the spin-one state; and $\delta_{ij}/\sqrt{N_c}$ corresponds to the color-singlet heavy quarkonium.  
In leading-order (LO) $v^2$ calculations, $q = 0$ is assumed. In subsequent calculations, we employ these spin projectors to incorporate the bound-state effect for both leptonium and quarkonium.

%%%%%%%%%%%%%%%%%%%%%%%%%%%%%%%%%
\subsection{$J/\psi \longrightarrow l^+ l^-$}
\label{subsec:jpsi2ll}
%%%%%%%%%%%%%%%%%%%%%%%%%%%%%%%%%

The decay width of the process $J/\psi \longrightarrow l^+l^-$ takes the simple form:
\begin{equation} \label{eq:jpsi2ll}
    \Gamma\left(J / \psi \rightarrow l^{+} l^{-}\right)=\frac{\alpha^2 e_Q^2 R_{J/\psi}^2(0)}{ m_c^2} \sqrt{1-\frac{m_l^2}{m_c^2}} \left(1+\frac{m_l^2}{2 m_c^2}\right),
\end{equation}
where $e_Q = 2/3$ denotes the electric charge of the charm quark, $m_l$ and $m_c$ are the lepton mass and charm quark mass, respectively, and $R_{J/\psi}(0)$ is the radial wavefunction of $J/\psi$ at the origin.

For the ratio of muon pair production to electron pair production, using Eq. \eqref{eq:jpsi2ll} and input parameters from PDG2024 \cite{ParticleDataGroup:2024cfk}, our theoretical prediction is:
\begin{equation}
    \left.\frac{\Gamma\left(J / \psi \rightarrow \mu^{+} \mu^{-}\right)}{\Gamma\left(J / \psi \rightarrow e^{+} e^{-}\right)}
    \right|_{\mathrm{the}}= 0.999992 \pm (6\times 10^{-11}),
\end{equation}
where the error, added in quadrature, originates from the three mass parameters. Using the branching fractions $Br\left(J / \psi \rightarrow \mu^{+} \mu^{-}\right) = (5.961 \pm 0.033)\%$ and $Br\left(J / \psi \rightarrow e^{+} e^{-} \right)= (5.971 \pm 0.032)\%$ from PDG2024, the experimental ratio is obtained as:
\begin{equation}
        \left.\frac{\Gamma\left(J / \psi \rightarrow \mu^{+} \mu^{-}\right)}{\Gamma\left(J / \psi \rightarrow e^{+} e^{-}\right)} \right|_{\mathrm{exp}}= 0.9983 \pm 0.0077.
\end{equation}
The theoretical prediction is consistent with the experimental result.

Using the experimental decay width $\Gamma\left(J / \psi \rightarrow \mu^{+} \mu^{-} \right)$, the radial wavefunction of $J/\psi$ at the origin can be extracted via Eq. \eqref{eq:jpsi2ll}:
\begin{equation}
    R_{J/\psi}^2(0) = 0.5599 \pm 0.0107 \,\, \mathrm{GeV}^3,
\end{equation}
which will be adopted in subsequent numerical evaluations.

%%%%%%%%%%%%%%%%%%%%%%%%%%%%%%%%%
\subsection{$J/\psi \longrightarrow l^+ l^- \gamma$}
\label{subsec:jpsi2lla}
%%%%%%%%%%%%%%%%%%%%%%%%%%%%%%%%%
\begin{figure}
    \centering
    \includegraphics[width=.8\linewidth]{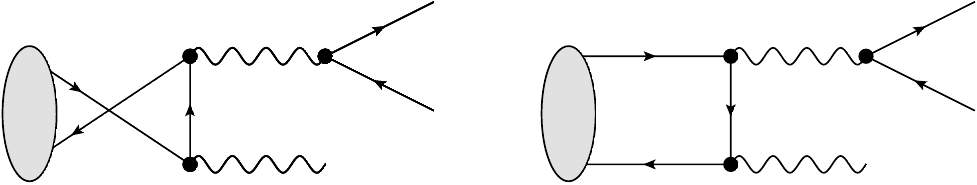}
    \includegraphics[width=.8\linewidth]{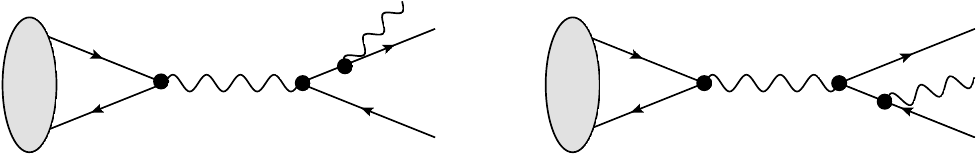}
    \caption{The Feynman diagrams for $J/\psi \longrightarrow l^+ l^- \gamma$ process.}
    \label{fig:jpsi2lla}
\end{figure}

The Feynman diagrams for $J/\psi$ decays into a lepton pair and a photon are shown in Fig. \ref{fig:jpsi2lla}. The first two diagrams (corresponding to $t$- and $u$-channels) are forbidden due to $C$-parity conservation, which has been verified via explicit calculation.

The differential decay width of $J/\psi(k) \longrightarrow l^+(p_1) l^-(p_2) \gamma(p_3)$ with respect to the invariant mass of the lepton pair, $s_1=(p_1+p_2)^2$, and that of the $l^-\gamma$ system, $s_2=(p_2+p_3)^2$, is given by:
\begin{equation} \label{eq:dGds1ds2}
    \frac{d\Gamma\left(J / \psi \rightarrow l^{+} l^{-} \gamma \right)}{ds_1 ds_2} = \frac{1}{3} \frac{1}{256\pi^3m^3_{J/\psi}} |\mathcal{M}|^2,
\end{equation}
where the upper and lower bounds for $s_1$ and $s_2$ can be found in the ``Kinematics'' section of PDG2024 \cite{ParticleDataGroup:2024cfk}.

It is of interest to examine the differential distribution $d\Gamma/ds_1$ with respect to the invariant mass of the lepton pair $s_1$, as the lepton pair may form a bound state (leptonium). This distribution is:
\begin{flalign} \label{eq:jpsi2lla_ds1}
    \frac{d\Gamma\left(J / \psi \rightarrow l^{+} l^{-} \gamma \right)}{ds_1} &= \frac{4 \alpha^3 e_Q^2 R_{J/\psi}^2(0) }{\pi m_{J/\psi}^6 (m_{J/\psi}^2 - s_1)}
    \times \left[  -\left(m_{J/\psi}^4 + s_1(s_1 +4 m_l^2) \right) \beta_l \right. \nonumber \\
 & \left. + \left( (m_{J/\psi}^2 -2 m_l^2)^2 + (s_1 +2 m_l^2)^2 -16 m_l^4  \right) \mathrm{log} \left(\frac{1 + \beta_l}{1 - \beta_l} \right) \right], 
\end{flalign}
where $\beta_l = \sqrt{1-4m_l^2/s_1}$. Under the approximation $m_l^2 \ll m^2_{J/\psi}$, Eq. \eqref{eq:jpsi2lla_ds1} simplifies to:
\begin{equation} \label{eq:jpsi2lla_ds1approxm}
    \left.\frac{d\Gamma\left(J / \psi \rightarrow l^{+} l^{-} \gamma \right)}{ds_1} \right|_{m^2_l \ll m^2_{J/\psi}} = \frac{4 \alpha^3 e_Q^2 R_{J/\psi}^2(0) \left(m_{J/\psi}^4 +s_1^2\right)}{\pi m_{J/\psi}^6 \left(m^2_{J/\psi} -s_1\right)} \left( \mathrm{log} \left(\frac{1 + \beta_l}{1 - \beta_l} \right) - \beta_l  \right).
\end{equation}
Eq. \eqref{eq:jpsi2lla_ds1approxm} serves as an excellent approximation for electrons and muons, with no visible distinction from Eq. \eqref{eq:jpsi2lla_ds1} in graphical representations. Further assuming $s_1$ lies in the leptonium threshold region ($s_1 \ll m^2_{J/\psi}$) under $m_l^2 \ll m^2_{J/\psi}$, Eq. \eqref{eq:jpsi2lla_ds1} simplifies to:
\begin{equation} \label{eq:jpsi2lla_ds1approx}
    \left.\frac{d\Gamma\left(J / \psi \rightarrow l^{+} l^{-} \gamma \right)}{ds_1} \right|_{m_l^2,s_1 \ll m^2_{J/\psi}} = \frac{4 \alpha^3 e_Q^2 R_{J/\psi}^2(0) }{\pi m_{J/\psi}^4} \left( \mathrm{log} \left(\frac{1 + \beta_l}{1 - \beta_l} \right) - \beta_l  \right).
\end{equation}

\begin{figure}
    \centering
    \includegraphics[width=1\linewidth]{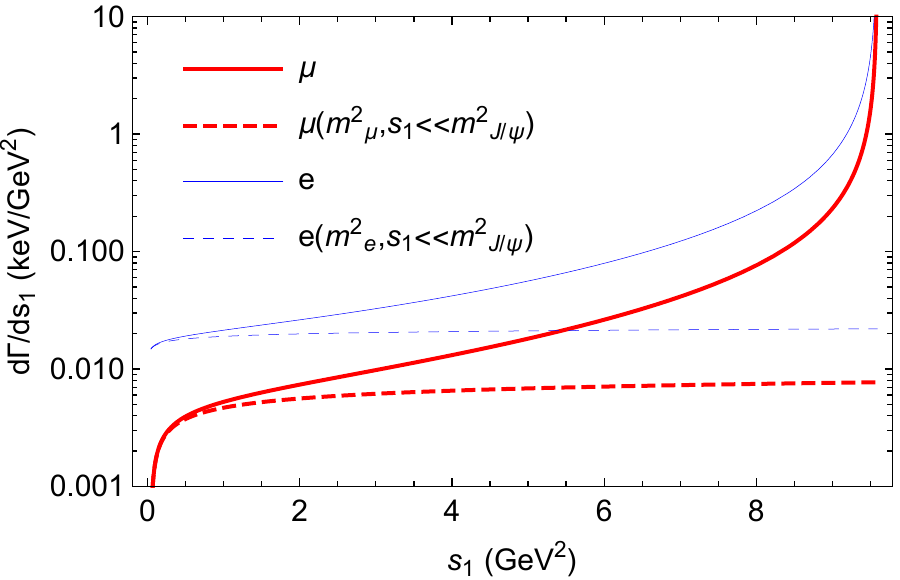}
    \caption{Differential distributions $d\Gamma\left(J / \psi \rightarrow l^{+} l^{-} \gamma \right)/ds_1$ for muons (thick lines) and electrons (thin lines). Solid lines correspond to the full formula in Eq. \eqref{eq:jpsi2lla_ds1}, and dashed lines to the approximation in Eq. \eqref{eq:jpsi2lla_ds1approx} under $m_l^2,s_1 \ll m^2_{J/\psi}$.}
    \label{fig:dGds1}
\end{figure}

In Fig. \ref{fig:dGds1}, we present the differential distributions $d\Gamma\left(J / \psi \rightarrow l^{+} l^{-} \gamma \right)/ds_1$. Thick lines correspond to $J / \psi \rightarrow \mu^{+} \mu^{-} \gamma$, while thin lines represent $J / \psi \rightarrow e^{+} e^{-} \gamma$. Solid lines are derived from the full formula in Eq. \eqref{eq:jpsi2lla_ds1}, and dashed lines from the approximate Eq. \eqref{eq:jpsi2lla_ds1approx} under the assumption $m_l^2,s_1 \ll m^2_{J/\psi}$. Eq. \eqref{eq:jpsi2lla_ds1approx} is a valid approximation only in the leptonium threshold region $s_1 \sim 4 m_l^2$, where the factor $ \mathrm{log} \left(\frac{1 + \beta_l}{1 - \beta_l} \right) - \beta_l$ behaves as $\sim \beta_l$. For a fixed $s_1$, $\beta_e > \beta_\mu$, leading to electron-channel curves lying above those for muons.

To compute the total decay widths, a minimum energy threshold for the emitted photon $E_\gamma$ is required to avoid infrared divergences from soft photons. This infrared divergence manifests in the factor $1/(m_{J/\psi}^2 - s_1)$ in Eq. \eqref{eq:jpsi2lla_ds1}. In the BESIII experiment, ``good photon'' selection imposes energy constraints: at least 25 MeV in the barrel region ($|\cos \theta| < 0.80$) and 50 MeV in the endcap region ($0.86 < |\cos \theta| < 0.92$), where $\theta$ denotes the polar angle of the photon track \cite{BESIII:2024jzg}. In our calculation, we adopt a photon energy cut $E_\gamma = \frac{m^2_{J/\psi} - s_1}{2 \sqrt{s_1}} > 100$ MeV, which yields a maximum $s_1$ of $s_{1\mathrm{max}} = 8.99 \, \mathrm{GeV}^2$. This choice facilitates comparison with the experimental branching fraction $Br(J/\psi \longrightarrow e^+e^- \gamma)\left.\right|_{\mathrm{exp}} = (0.88 \pm 0.14) \%$ \cite{ParticleDataGroup:2024cfk}.

The total decay widths for $J/\psi$ decays into a lepton pair and a photon with the cut $E_\gamma > 100$ MeV are:
\begin{align}
    \Gamma\left(J / \psi \rightarrow \mu^{+} \mu^{-} \gamma \right) &= 0.283 \pm 0.005\, \mathrm{keV}, \\
    \Gamma\left(J / \psi \rightarrow e^{+} e^{-} \gamma \right) &= 0.854 \pm 0.016 \, \mathrm{keV}.
\end{align}
Using the total decay width of $J/\psi$, $\Gamma_{J/\psi} = 92.6 \pm 1.7$ keV from PDG2024, the corresponding branching fractions are:
\begin{align}
    Br\left(J / \psi \rightarrow \mu^{+} \mu^{-} \gamma \right) &= (0.31 \pm 0.01) \% , \\
    Br\left(J / \psi \rightarrow e^{+} e^{-} \gamma \right) &= (0.92 \pm 0.02 ) \%.
\end{align}
The branching fraction for $J/\psi \rightarrow e^{+} e^{-} \gamma$ is consistent with the experimental result $Br(J/\psi \longrightarrow e^+e^- \gamma)\left.\right|_{\mathrm{exp}} = (0.88 \pm 0.14) \%$ \cite{ParticleDataGroup:2024cfk}. No experimental measurement exists for the branching fraction of $J/\psi \rightarrow \mu^{+} \mu^{-} \gamma$ to date. Although both $m_e^2$ and $m_\mu^2$ are much smaller than $m_{J/\psi}^2$, the mass hierarchy $m_e^2 \ll m_\mu^2$ results in a significant difference between muon and electron channels in $J/\psi \longrightarrow l^+l^- \gamma$ processes:
\begin{equation}
    \frac{Br\left(J / \psi \rightarrow \mu^{+} \mu^{-} \gamma \right)}{Br\left(J / \psi \rightarrow e^{+} e^{-} \gamma \right)} = 0.33 \pm 0.01.
\end{equation}

%%%%%%%%%%%%%%%%%%%%%%%%%%%%%%%%%
\subsection{$J/\psi \longrightarrow (l^+ l^-)[^1S_0] + \gamma$}
\label{subsec:jpsi2leptoniuma}
%%%%%%%%%%%%%%%%%%%%%%%%%%%%%%%%%

Since the $J/\psi$ (with $J^{PC}=1^{--}$) can only decay into leptonium $(l^+ l^-)$ and a photon via a single virtual photon, as depicted in the last two Feynman diagrams of Fig. \ref{fig:jpsi2lla}, the $J^{PC}$ quantum number of the leptonium must be $0^{-+}$ (i.e., the $^1S_0$ state) due to $C$-parity conservation.  
The squared amplitude takes the simple form:
\begin{equation}
    \left| \cal{M} \right|^2 = \frac{512 \pi e_Q^2 \alpha^3 R_{J/\psi}(0)^2 R_{L}(0)^2}{m_L m_{J/\psi}^3},
\end{equation}
where $m_L$ denotes the leptonium mass and $R_{L}(0)$ is the radial wavefunction of the leptonium at the origin. The corresponding decay width is given by:
\begin{equation} \label{eq:jpsi2La}
    \Gamma(J/\psi \longrightarrow (l^+l^-)[^1S_0] + \gamma) = \frac{32 e_Q^2 \alpha^3 R_{J/\psi}(0)^2 R_{L}(0)^2 \left(m_{J/\psi}^2 - m_L^2 \right)}{m_L m_{J/\psi}^6}.
\end{equation}
Numerically, using the total decay width $\Gamma_{J/\psi} = 92.6 \pm 1.7$ keV, the resulting branching fractions are:
\begin{align}
     Br(J/\psi \longrightarrow (\mu^+\mu^-)[^1S_0] + \gamma) &= (3.93 \pm 0.13) \times 10^{-13}, \label{eq:jps2La-1}  \\
     Br(J/\psi \longrightarrow (e^+e^-)[^1S_0] + \gamma) &= (9.23 \pm 0.31) \times 10^{-18}.  \label{eq:jps2La-2}
\end{align}
These branching fractions are extremely small due to the $\alpha^6 \approx 10^{-13}$ suppression factor in the decay width, where $\alpha^3$ originates from the leptonium wavefunction at the origin. Although $10^{10}$ $J/\psi$ events have been accumulated at BESIII \cite{BESIII:2021cxx}, no signals of dimuonium or positronium from $J/\psi \longrightarrow (l^+ l^-)[^1S_0] + \gamma$ decays can be detected. However, with the annual production of $3.4 \times 10^{12}$ $J/\psi$ events at the future STCF \cite{Achasov:2023gey}, long-term accumulation could make the observation of dimuonium $(\mu^+\mu^-)[^1S_0]$ via such radiative $J/\psi$ decays promising.

To compare with Ref. \cite{Dai:2024yhy}, where dimuonium production via $J/\psi$ radiative decay is studied using a method in which the bound state of lepton pairs is described by the Green function, we define the ratio  
\begin{equation}
    R_l= \frac{\Gamma(J/\psi \longrightarrow (l^+l^-)[^1S_0] + \gamma)}{\Gamma\left(J / \psi \rightarrow l^{+} l^{-}\right)}.
\end{equation}  
By taking $m_L =2 m_l$, $m_{J/\psi} =2 m_c$, and $R_{L}(0)^2 = \frac{(\alpha m_l)^3}{2n^3}$ (with $n=1$ for the ground state), this ratio simplifies to  
\begin{align}
    R_l &= \frac{\alpha^4 m_l^2 \sqrt{1- m_l^2/m_c^2}}{2m_c^2 + m_l^2} \label{eq:R1} \\
    &\approx \frac{\alpha^4 }{2} \left(\frac{m_l}{m_c}\right)^2, \label{eq:R}
\end{align}  
where the second line is derived under the assumption $m_l^2 \ll m_c^2$.  
Notably, Eq. \eqref{eq:R1} differs from the first line of Eq. (32) in Ref. \cite{Dai:2024yhy} by a factor of $\left(1- m_l^2/m_c^2 \right) \zeta(3)$. 
The factor $\left(1- m_l^2/m_c^2 \right)$ originates from the $1 \to 2$ phase space of the decay $J/\psi \rightarrow (l^+l^-)[^1S_0] + \gamma$. Specifically, in Eq. (24) of Ref. \cite{Dai:2024yhy}, the differential decay width for the target $1 \to 2$ process $J/\psi \rightarrow (\mu^+\mu^-) + \gamma$ is obtained via the Green function method from the $1 \to 3$ process $J/\psi \rightarrow \mu^+\mu^- \gamma$, leading to the discrepancy introduced by $\left(1- m_l^2/m_c^2 \right)$.  
In the limit $m_l^2 \ll m_c^2$, this factor approaches 1, and Eq. \eqref{eq:R} then differs from the second line of Eq. (32) in Ref. \cite{Dai:2024yhy} only by a factor of $\zeta(3) = 1.20$. This factor arises from summing over all higher excited states of leptonium, i.e., $\sum_{n=1}^\infty \frac{1}{n^3} = \zeta(3)$, whereas we restrict our analysis to the ground state ($n=1$) here. Had we included this contribution by adopting $R_{L}(0)^2 = \frac{(\alpha m_l)^3}{2 n^3}$ and summing over $n$, our result would agree with that of Ref. \cite{Dai:2024yhy} in the limit $m_l^2 \ll m_c^2$. This indicates that contributions from higher excited states of leptonium account for approximately 20\%. This conclusion applies to all leptonium-related calculations in this paper.  
Eq. \eqref{eq:R} also implies that  
\begin{equation}
    \frac{R_\mu}{R_e} = \frac{\Gamma(J/\psi \longrightarrow (\mu^+\mu^-)[^1S_0] + \gamma)}{\Gamma(J/\psi \longrightarrow (e^+e^-)[^1S_0] + \gamma)} \approx \frac{m_\mu^2}{m_e^2},
\end{equation}  
which explicitly reflects the mass hierarchy between muons and electrons.

%%%%%%%%%%%%%%%%%%%%%%%%%%%%%%%%%
\subsection{$J/\psi \longrightarrow (l_1^+ l_2^-)[n] + l_1^-l_2^+$}
\label{subsec:jpsi2leptoniumall}
%%%%%%%%%%%%%%%%%%%%%%%%%%%%%%%%%
\begin{figure}
    \centering
    \includegraphics[width=0.9\linewidth]{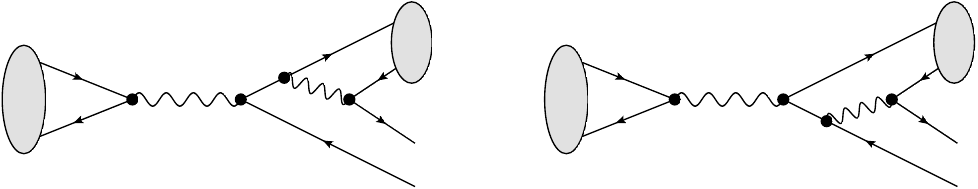}
    \includegraphics[width=0.9\linewidth]{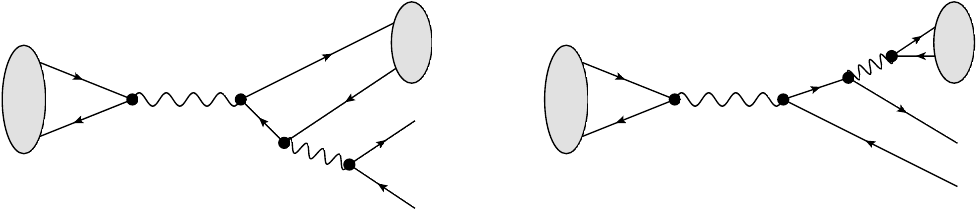}
    \caption{The typical Feynman diagrams for $J/\psi \longrightarrow (l_1^+ l_2^-)[n] + l_1^-l_2^+$ ($l_{1,2} = e,\mu$) process. There are four additional diagrams by exchanging the two fermion lines in the final states.}
    \label{fig:jpsi2leptoniumll}
\end{figure}

The Feynman diagrams for the $J/\psi \longrightarrow (l_1^+ l_2^-)[n] + l_1^-l_2^+$ process are displayed in Fig. \ref{fig:jpsi2leptoniumll}. Here, $l_{1,2}$ can be either muons or electrons, and $n$ corresponds to either the $^1S_0$ or $^3S_1$ state. Four additional diagrams, where the two fermion lines in the final state are exchanged, are implicitly included in the following discussion. Notably, the 3rd diagram vanishes for $^3S_1$ states, and the 4th diagram vanishes for $^1S_0$ states, due to $C$-parity conservation.

For $J/\psi \longrightarrow (\mu^+ e^-)[n] + \mu^- e^+$, the top two diagrams in Fig. \ref{fig:jpsi2leptoniumll} contribute. The explicit squared amplitudes for $J/\psi \longrightarrow (\mu^+ e^-)[n] (p_1) + e^+(p_2)+ \mu^-(p_3)$ are lengthy, but they simplify in the limit $m_e^2 \ll m_{J/\psi}^2$:
\begin{align} \label{eq:sqM1S0}
    \left| \cal{M} \right|^2 ((\mu^+ e^-)[^1S_0]) &\approx \frac{ 1024 \pi^2 \alpha^7 m_\mu^2 R_{J/\psi}(0)^2}{9 m_{J/\psi}^3 \left(s_2 -m_\mu^2 \right)^2 \left(s_1 -m_\mu^2 \right)^3} \left[ m_{J/\psi}^4 \left(s_1 -m_\mu^2 \right) \left(s_2 -m_\mu^2 \right) \right. \nonumber \\
    & \left. - m_{J/\psi}^2 \left(2 m_\mu^4 s_1 + m_\mu^2 (s_2^2 -s_1^2 -4s_1 s_2) +2 s_1^2 s_2 \right) - 3 m_\mu^8 +3 m_\mu^6 (s_1 -s_2) \right. \nonumber \\
    & \left. - m_\mu^4 (s_1^2 -7 s_1 s_2 -4 s_2^2) - m_\mu^2 s_2 (4s_1^2 +3s_1s_2 +s_2^2) + s_1^3s_2 \right],
\end{align}
\begin{align} \label{eq:sqM3S1}
    \left| \cal{M} \right|^2 ((\mu^+ e^-)[^3S_1]) &\approx \frac{ 1024 \pi^2 \alpha^7 m_\mu^2 R_{J/\psi}(0)^2}{9 m_{J/\psi}^3 \left(s_2 -m_\mu^2 \right)^2 \left(s_1 -m_\mu^2 \right)^3} \left[ 3 m_{J/\psi}^4 \left(s_1 -m_\mu^2 \right) \left(s_2 -m_\mu^2 \right) \right. \nonumber \\
    & \left. - m_{J/\psi}^2 \left(2 m_\mu^4 (s_1 +2s_2) + m_\mu^2 (s_1^2 -s_2^2 -12s_1 s_2) +2 s_1 s_2 (s_1 +2s_2) \right) \right. \nonumber \\
    & \left. - 9 m_\mu^8 - 3 m_\mu^6 (s_1 -s_2) + m_\mu^4 (7s_1^2 +21 s_1 s_2 +2 s_2^2) \right. \nonumber \\ 
    & \left. - m_\mu^2 (2s_1^3 + 10 s_1^2 s_2 + 11s_1s_2^2 +s_2^3) + s_1 s_2(s_1^2 +2 s_2^2) \right],
\end{align}
where $s_1 = (p_1 +p_2)^2$ and $s_2=(p_2+p_3)^2$. Retaining only the leading terms proportional to $m_{J/\psi}^4$ and $m_\mu^8$ within the square brackets, the squared amplitude for the $^3S_1$ state in Eq. \eqref{eq:sqM3S1} is three times that of the $^1S_0$ state in Eq. \eqref{eq:sqM1S0}.

Integrating over the 3-body phase space (as defined in Eq. \eqref{eq:dGds1ds2}), we obtain the total decay widths:
\begin{align}
    \Gamma(J/\psi \longrightarrow (\mu^+ e^-)[^1S_0] + e^+\mu^-) &= 4.90 \times 10^{-10} \, \mathrm{keV},\\
    \Gamma(J/\psi \longrightarrow (\mu^+ e^-)[^3S_1] + e^+\mu^-) &= 1.40 \times 10^{-9} \, \mathrm{keV},
\end{align}
where the decay width for the $(\mu^+ e^-)[^3S_1]$ state is approximately three times that of the $(\mu^+ e^-)[^1S_0]$ state, as expected. Using the $J/\psi$ total width $\Gamma_{J/\psi} = 92.6$ keV, the corresponding branching fractions are:
\begin{align}
    Br(J/\psi \longrightarrow (\mu^+ e^-)[^1S_0] + e^+\mu^-) &= 5.3 \times 10^{-12},\\
    Br(J/\psi \longrightarrow (\mu^+ e^-)[^3S_1] + e^+\mu^-) &= 1.5 \times 10^{-11}.
\end{align}
With the annual production of $3.4 \times 10^{12}$ $J/\psi$ events at the future STCF, the observation of muonium $(\mu^+ e^-)[n]$ within several years appears feasible.

\begin{figure}
    \centering
    \includegraphics[width=0.32\linewidth]{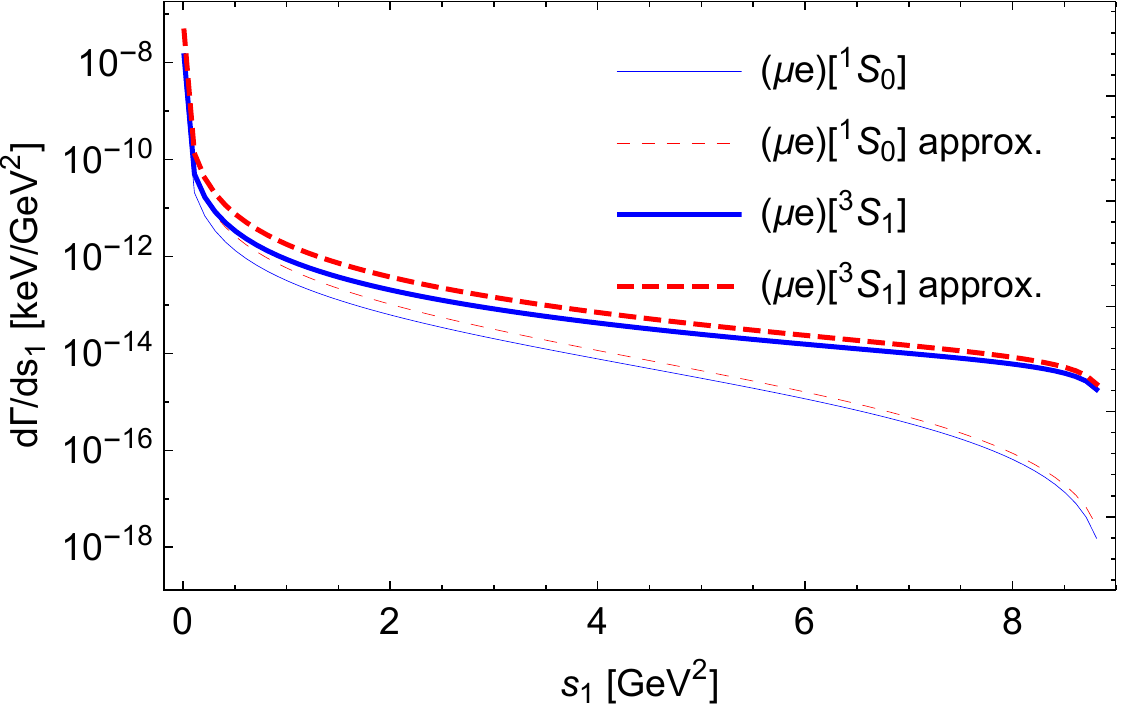}
    \includegraphics[width=0.32\linewidth]{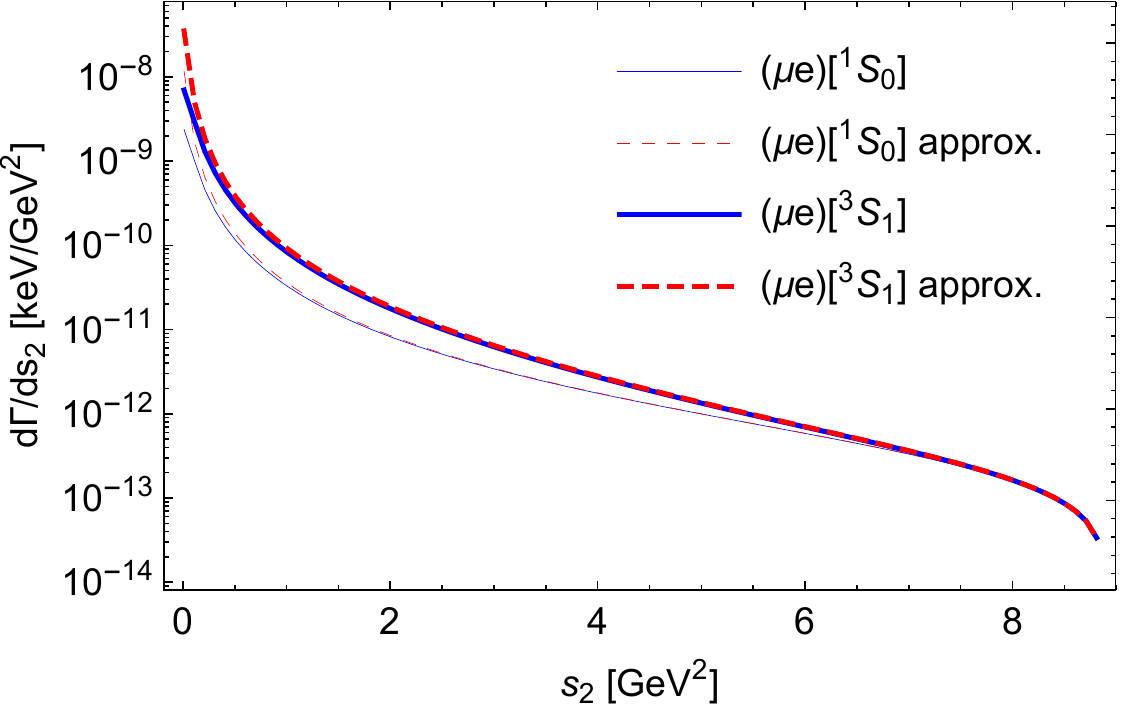}
    \includegraphics[width=0.32\linewidth]{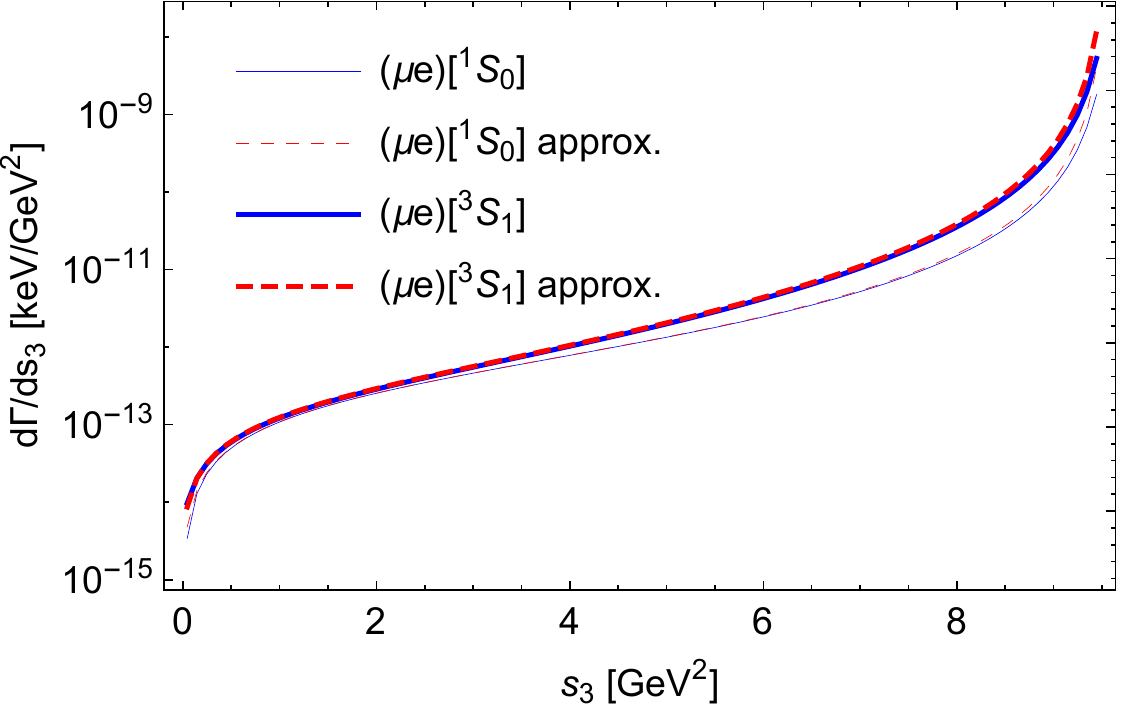}
    \caption{Differential distributions for $J/\psi \longrightarrow (\mu^+ e^-)[n] + e^++ \mu^-$ versus the invariants $s_1$, $s_2$, and $s_3$. Curves labeled ``approx.'' correspond to the $m_e^2 \ll m_{J/\psi}^2$ limit.}
    \label{fig:dGds123ue}
\end{figure}

In Fig. \ref{fig:dGds123ue}, we present the differential distributions for $J/\psi \longrightarrow (\mu^+ e^-)[n] (p_1) + e^+(p_2) + \mu^-(p_3)$ versus the invariants $s_1 = (p_1 + p_2)^2$, $s_2 = (p_2+p_3)^2$, and $s_3=(p_1+p_3)^2$, where $s_3 = m_{J/\psi}^2 + (m_\mu +m_e)^2 +m_\mu^2 +m_e^2- s_1 -s_2$. The approximated curves correspond to Eqs. \eqref{eq:sqM1S0} and \eqref{eq:sqM3S1} in the $m_e^2 \ll m_{J/\psi}^2$ limit. We find that $d\Gamma/ds_1$ and $d\Gamma/ds_2$ exhibit similar behaviors, peaking in low $s_{1,2}$ regions, while $d\Gamma/ds_3$ peaks in the high $s_3$ region. This behavior arises because the mass of $e^+ (p_2)$ is much smaller than those of $(\mu^+ e^-)[n] (p_1)$ and $\mu^-(p_3)$, and is further influenced by the $1/\left[\left(s_2 -m_\mu^2 \right)^2 \left(s_1 -m_\mu^2 \right)^3 \right]$ factors in the approximated squared amplitudes (Eqs. \eqref{eq:sqM1S0} and \eqref{eq:sqM3S1}).

For $J/\psi \longrightarrow (l_1^+ l_1^-)[n](p_1) + l_2^+(p_2) + l_2^-(p_3) $ ($l_{1,2} = \mu ,e$, with $l_1 \neq l_2$), the 3rd and 4th diagrams in Fig. \ref{fig:jpsi2leptoniumll} contribute, with the 3rd diagram vanishing for $^3S_1$ states and the 4th diagram vanishing for $^1S_0$ states. In the limit $m_e^2 \ll m_{J/\psi}^2$, the squared amplitudes for $J/\psi \longrightarrow (\mu^+ \mu^-)[n]+e^+e^-$ are:
\begin{align} \label{eq:sqM1S0diuee}
    \left| \cal{M} \right|^2 ((\mu^+ \mu^-)[^1S_0]+e^+e^-) &\approx \frac{ 1024 \pi^2 e_Q^2  \alpha^7 R_{J/\psi}(0)^2}{ m_{J/\psi} s_2 \left( m_{J/\psi}^2 +s_2 -4m_\mu^2 \right)^2} \left( \frac{m_\mu}{m_{J/\psi}} \right)^2  \nonumber \\
    & \times \left[ \left(m_{J/\psi}^2 + 4 m_\mu^2 -s_1-s_2 \right)^2 - 8 m_{J/\psi}^2 m_\mu^2 +s_1^2  \right],
\end{align}
\begin{align} \label{eq:sqM3S1diuee}
    \left| \cal{M} \right|^2 ((\mu^+ \mu^-)[^3S_1]+e^+e^-) &\approx 
    \frac{128 \pi^2 e_Q^2 \alpha^7 R_{J/\psi}(0)^2 }{ m_{J/\psi}^3 s_1^2 \left( m_{J/\psi}^2 + 4m_\mu^2 -s_1 -s_2 \right)^2 } \nonumber \\
    & \hspace{-4cm} \times  \left[ m_{J/\psi}^6 \left(s_1 - 4m_\mu^2\right) 
    - m_{J/\psi}^4 \left( 32m_\mu^4 -4 m_\mu^2 (5s_1 +2 s_2) + s_1(3s_1+s_2)\right) \right. \nonumber \\ 
    & \hspace{-4cm} \left. + m_{J/\psi}^2 \left( -64 m_\mu^6 + 16 m_\mu^4(5s_1 +2s_2) -4m_\mu^2(8s_1^2 +4s_1s_2 +s_2^2) +s_1(2s_1+s_2)^2 \right) \right. \nonumber \\  
    & \hspace{-4cm} \left. + s_1 \left( 4m_\mu^2 -s_1 -s_2 \right) \left( 16 m_\mu^4 - 8 m_\mu^2s_1 +2s_1^2 +2s_1s_2 +s_2^2 \right)  \right].
\end{align}
For $J/\psi \longrightarrow (e^+e^-)[n]+\mu^+ \mu^-$ in the $m_e^2 \ll m_{J/\psi}^2$ limit, the squared amplitudes are:
\begin{align} \label{eq:sqM1S0dieuu}
    \left| \cal{M} \right|^2 ((e^+e^-)[^1S_0]+\mu^+ \mu^-) &\approx \frac{ 1024 \pi^2 e_Q^2 \alpha^7 R_{J/\psi}(0)^2}{ m_{J/\psi} s_2^2 \left( m_{J/\psi}^2 +s_2 \right)^2} \left( \frac{m_e}{m_{J/\psi}} \right)^2 
 \nonumber \\
    & \hspace{-4cm} \times \left[ m_{J/\psi}^4 \left(2m_\mu^2 + s_2\right) -2m_{J/\psi}^2 s_2 \left( m_\mu^2 +s_1+s_2 \right) + s_2 \left(2(m_\mu^2 -s_1)^2 +2s_1s_2 +s_2^2\right) \right],
\end{align}
\begin{align} \label{eq:sqM3S1dieuu}
    \left| \cal{M} \right|^2 ((e^+e^-)[^3S_1]+\mu^+ \mu^-) &\approx 
    \frac{128 \pi^2 e_Q^2 \alpha^7 R_{J/\psi}(0)^2 }{ m_{J/\psi}^3 \left(m_\mu^2 - s_1\right)^2 \left( m_{J/\psi}^2 + m_\mu^2 -s_1 -s_2 \right)^2 } \nonumber \\
    & \hspace{-4cm} \times  \left[ m_{J/\psi}^6 \left(s_1 - 3m_\mu^2\right) 
    - m_{J/\psi}^4 \left( 7m_\mu^4 - m_\mu^2 (6s_1 +5 s_2) + s_1(3s_1+s_2)\right) \right. \nonumber \\ 
    & \hspace{-4cm} \left. + m_{J/\psi}^2 \left( 4 s_2 (2 m_\mu^4 -m_\mu^2s_1+s_1^2) +s_2^2(s_1 -3m_\mu^2) -4(m_\mu^2-s_1)^3\right) \right. \nonumber \\  
    & \hspace{-4cm} \left. + s_2^2 \left( 2m_\mu^2s_1 -3m_\mu^4 -3s_1^2 \right) + s_2^3\left( m_\mu^2 -s_1 \right) -4 s_1 s_2\left(  m_\mu^2 -s_1\right)^2 -2 \left( m_\mu^2 -s_1\right)^4 \right].
\end{align}
The squared amplitudes for $^3S_1$ states differ significantly from those for $^1S_0$ states due to their distinct Feynman diagram topologies. This indicates that the 3rd diagram in Fig. \ref{fig:jpsi2leptoniumll} contributes at leading order $(\frac{m_l}{m_{J/\psi}})^2$, while the 4th diagram contributes at leading order $(\frac{m_l}{m_{J/\psi}})^0$. 

Integrating over the 3-body phase space, we obtain the total decay widths:
\begin{align}
    \Gamma(J/\psi \longrightarrow (\mu^+ \mu^-)[^1S_0] + e^+e^-) &= 3.81 \times 10^{-13} \, \mathrm{keV}, \label{eq:jpsi2Lmue-1}\\
    \Gamma(J/\psi \longrightarrow (\mu^+ \mu^-)[^3S_1] + e^+e^-) &= 6.70 \times 10^{-11} \, \mathrm{keV}; \label{eq:jpsi2Lmue-2}\\
    \Gamma(J/\psi \longrightarrow (e^+ e^-)[^1S_0] + \mu^+\mu^-) &= 1.92 \times 10^{-18} \, \mathrm{keV}, \label{eq:jpsi2Lemu-1}\\
    \Gamma(J/\psi \longrightarrow (e^+ e^-)[^3S_1] + \mu^+\mu^-) &= 5.68 \times 10^{-10} \, \mathrm{keV}. \label{eq:jpsi2Lemu-2}
\end{align}
The decay width for $J/\psi \longrightarrow (e^+ e^-)[^1S_0] + \mu^+\mu^-$ is extremely small due to the negligible contribution of the 3rd diagram, while the decay width for $J/\psi \longrightarrow (e^+ e^-)[^3S_1] + \mu^+\mu^-$ is significantly larger due to the dominant contribution of the 4th diagram. This trend holds for dimuonium but is less dramatic than for positronium. With the $J/\psi$ width $\Gamma_{J/\psi} = 92.6$ keV and the annual production of $3.4 \times 10^{12}$ $J/\psi$ events at STCF, the observation of $(e^+ e^-)[^3S_1]$ and even $(\mu^+ \mu^-)[^3S_1]$ via these channels within several years is plausible.

\begin{figure}
    \centering
    \includegraphics[width=0.32\linewidth]{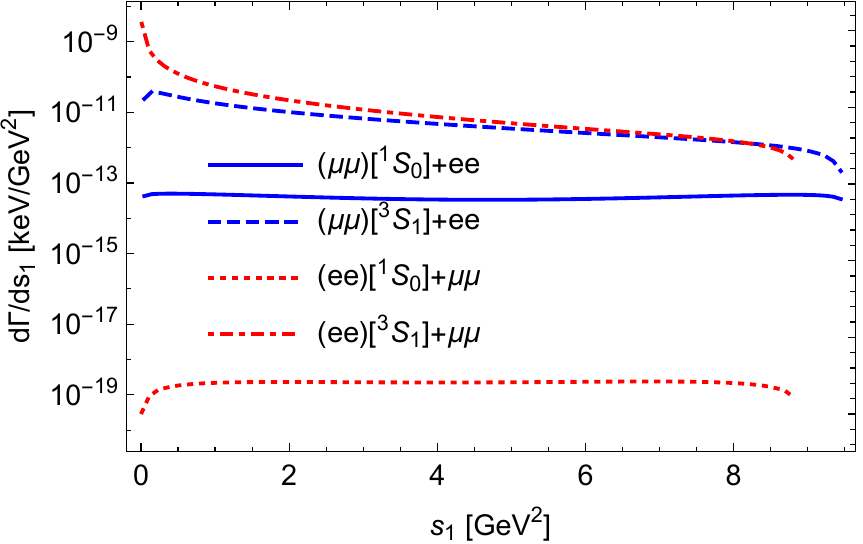}
    \includegraphics[width=0.32\linewidth]{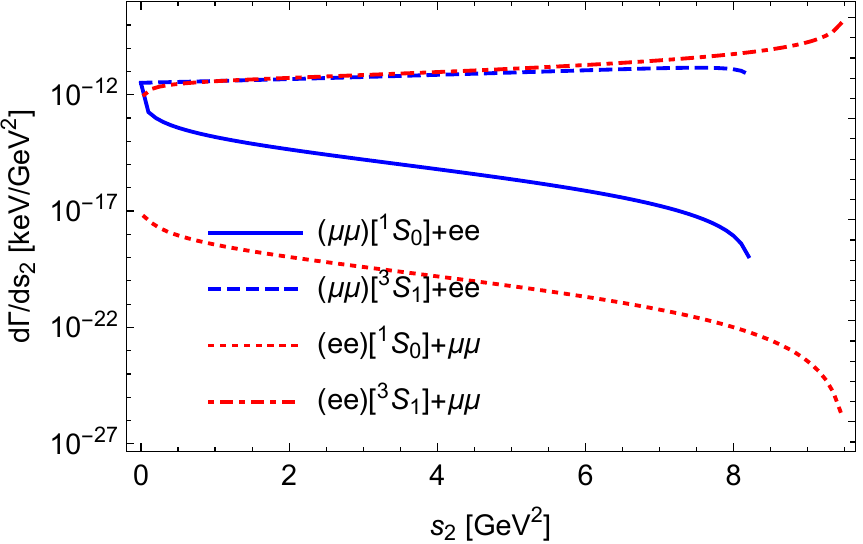}
    \includegraphics[width=0.32\linewidth]{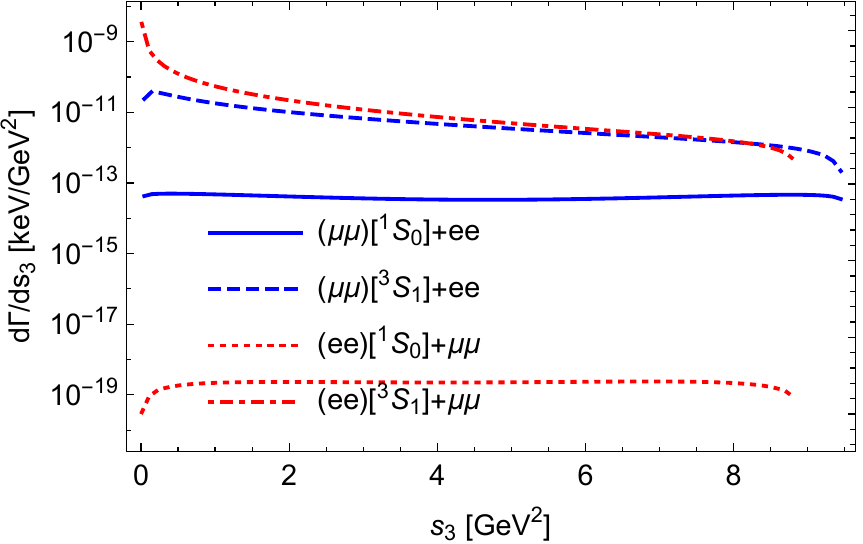}
    \caption{Differential distributions for $J/\psi \longrightarrow (l_1^+ l_1^-)[n] + l_2^+l_2^- $ ($l_{1,2} = \mu ,e$, $l_1 \neq l_2$) versus the invariants $s_1$, $s_2$, and $s_3$.}
    \label{fig:dGds123l1l2}
\end{figure}

In Fig. \ref{fig:dGds123l1l2}, we present the differential distributions $d\Gamma/ds_{1,2,3}$ for $J/\psi \longrightarrow (l_1^+ l_1^-)[n] + l_2^+l_2^- $ ($l_{1,2} = \mu ,e$, $l_1 \neq l_2$), comparing $J/\psi \longrightarrow (\mu^+ \mu^-)[n]+e^+e^-$ and $J/\psi \longrightarrow (e^+e^-)[n]+\mu^+ \mu^-$. States with the same spin exhibit similar distribution trends. The distributions $d\Gamma/ds_1$ and $d\Gamma/ds_3$ are identical because $l_2^{+}(p_2)$ and $l_2^{-}(p_3)$ are the same lepton type with opposite charges. The approximated amplitudes in Eqs. \eqref{eq:sqM1S0diuee}–\eqref{eq:sqM3S1dieuu} are in excellent agreement with the full results and are not explicitly shown.

For $J/\psi \longrightarrow (l^+ l^-)[n] (p_1) + l^+(p_2)+ l^-(p_3)$ with $l=\mu,\,e$, all four Feynman diagrams in Fig. \ref{fig:jpsi2leptoniumll} contribute, though the 3rd diagram vanishes for $^3S_1$ states and the 4th diagram vanishes for $^1S_0$ states. In the limit $m_l^2 \ll m_{J/\psi}^2$, the squared amplitudes are:
\begin{align} \label{eq:sqM1S0l}
    \left| \cal{M} \right|^2 ((l^+ l^-)[^1S_0]) &\approx \frac{ 256 \pi^2 e_Q^2 \alpha^7 R_{J/\psi}(0)^2}{ m_{J/\psi} s_1^2 s_2 \left(s_2 + m_{J/\psi}^2 \right)^2 \left( m_{J/\psi}^2- s_1 - s_2 \right)^2} \left( \frac{m_l}{m_{J/\psi}} \right)^2 \nonumber \\
    & \times \left[ m_{J/\psi}^8 \left(4 s_1^2 +14 s_1s_2 +9s_2^2 \right) - 2 m_{J/\psi}^6 \left(8s_1^3 + 39 s_1^2s_2 + 40 s_1 s_2^2 +12 s_2^3 \right) \right. \nonumber \\
    & \left. + 2m_{J/\psi}^4 \left(14 s_1^4 +72s_1^3 s_2 +97 s_1^2 s_2^2 +54 s_1 s_2^3 +11s_2^4 \right) \right. \nonumber \\ 
    & \left. - 2 m_{J/\psi}^2 \left( 12 s_1^5 +52s_1^4s_2 + 80s_1^3s_2^2 + 61 s_1^2 s_2^3 +24 s_1 s_2^4 +4s_2^5 \right) \right. \nonumber \\  
    & \left. + \left(2s_1^2 +2s_1s_2 +s_2^2 \right)^3  \right],
\end{align}
\begin{align} \label{eq:sqM3S1l}
    \left| \cal{M} \right|^2 ((l^+ l^-)[^3S_1]) &\approx 
    \frac{128 \pi^2 e_Q^2 \alpha^7 R_{J/\psi}(0)^2 }{ m_{J/\psi}^3 s_1 \left( m_{J/\psi}^2 -s_1 -s_2 \right) } \left(  (m_{J/\psi}^2 -s_1)^2 +(s_1 +s_2)^2 \right)\nonumber \\ 
    & + \frac{ 128 \pi^2 e_Q^2 \alpha^7 R_{J/\psi}(0)^2}{ m_{J/\psi} s_1^2 \left(s_2 + m_{J/\psi}^2 \right)^2 \left( m_{J/\psi}^2- s_1 - s_2 \right)^2} \left( \frac{m_l}{m_{J/\psi}} \right)^2 \nonumber \\
    & \times \left[ 11 m_{J/\psi}^{10} - m_{J/\psi}^8 \left(26 s_1 +5 s_2 \right) - 4 m_{J/\psi}^6 \left(7s_1^2 + 10 s_1s_2 + 19 s_2^2 \right) \right. \nonumber \\
    & \left. + 4 m_{J/\psi}^4 \left(24s_1^3 +28s_1^2 s_2 +44 s_1 s_2^2 +45 s_2^3 \right) \right. \nonumber \\ 
    & \left. - m_{J/\psi}^2 \left( 48 s_1^4 +32s_1^3s_2 + 124s_1^2s_2^2 + 112 s_1 s_2^3 +127 s_2^4 \right) \right. \nonumber \\  
    & \left. - s_2 \left(32s_1^4 + 64 s_1^3s_2 + 72s_1^2 s_2^2 +46s_1s_2^3 -17s_2^4 \right)  \right],
\end{align}
where the squared amplitude for the $^1S_0$ state starts at order $(\frac{m_l}{m_{J/\psi}})^2$, while that for the $^3S_1$ state starts at order $(\frac{m_l}{m_{J/\psi}})^0$. Numerical evaluations confirm that the zero-order term dominates for the $^3S_1$ state.

Integrating over the 3-body phase space, the total decay widths are:
\begin{align}
    \Gamma(J/\psi \longrightarrow (\mu^+ \mu^-)[^1S_0] + \mu^+\mu^-) &= 2.02 \times 10^{-12} \, \mathrm{keV}, \label{eq:jpsi2Lmumu-1}\\
    \Gamma(J/\psi \longrightarrow (\mu^+ \mu^-)[^3S_1] + \mu^+\mu^-) &= 7.11 \times 10^{-11} \, \mathrm{keV}; \label{eq:jpsi2Lmumu-2}\\
    \Gamma(J/\psi \longrightarrow (e^+ e^-)[^1S_0] + e^+e^-) &= 2.19 \times 10^{-12} \, \mathrm{keV}, \label{eq:jpsi2Lee-1}\\
    \Gamma(J/\psi \longrightarrow (e^+ e^-)[^3S_1] + e^+e^-) &= 9.66 \times 10^{-10} \, \mathrm{keV}. \label{eq:jpsi2Lee-2}
\end{align}
Decay widths for $^3S_1$ states are larger than those for $^1S_0$ states. Notably, the decay width of $(e^+ e^-)[^1S_0]$ is comparable to that of $(\mu^+ \mu^-)[^1S_0]$, while the decay width of $(e^+ e^-)[^3S_1]$ is approximately an order of magnitude larger than that of $(\mu^+ \mu^-)[^3S_1]$. With the $J/\psi$ width $\Gamma_{J/\psi} = 92.6$ keV and STCF's annual production of $3.4 \times 10^{12}$ $J/\psi$ events, the observation of $(e^+ e^-)[^3S_1]$ and even $(\mu^+ \mu^-)[^3S_1]$ within several years is promising.

\begin{figure}
    \centering
    \includegraphics[width=0.32\linewidth]{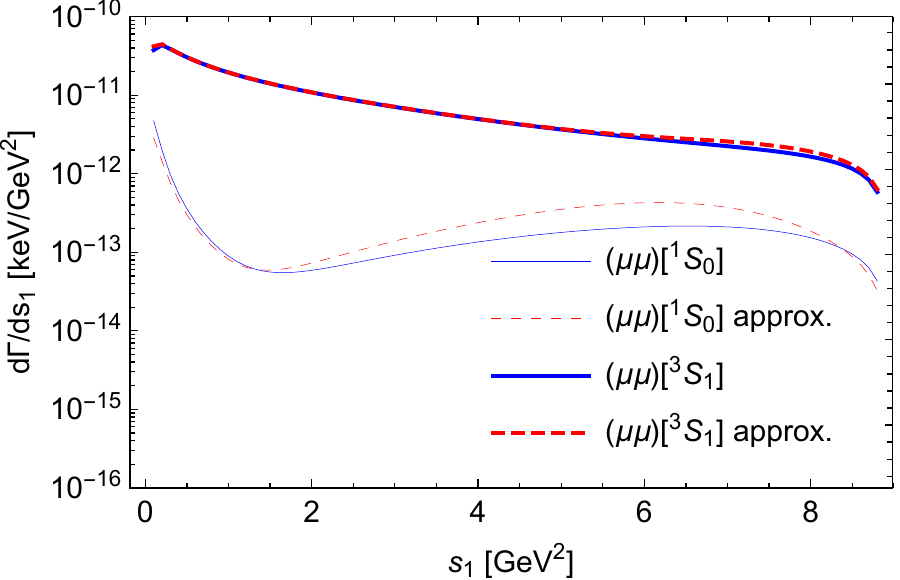}
    \includegraphics[width=0.32\linewidth]{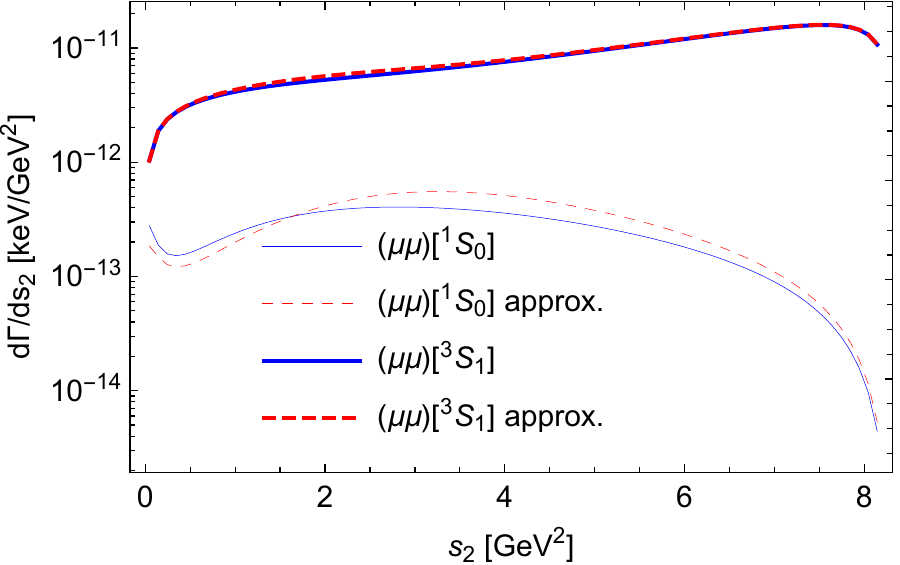}
    \includegraphics[width=0.32\linewidth]{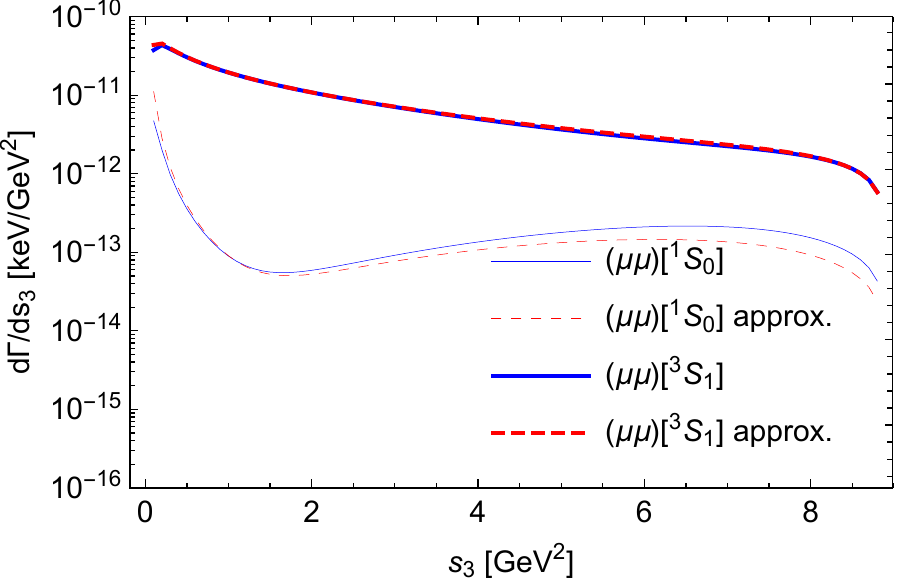}
    \caption{Differential distributions for $J/\psi \longrightarrow (\mu^+ \mu^-)[n] + \mu^++ \mu^-$ versus the invariants $s_1$, $s_2$, and $s_3$. Curves labeled ``approx.'' correspond to the $m_\mu^2 \ll m_{J/\psi}^2$ limit.}
    \label{fig:dGds123uu}
\end{figure}

\begin{figure}
    \centering
    \includegraphics[width=0.32\linewidth]{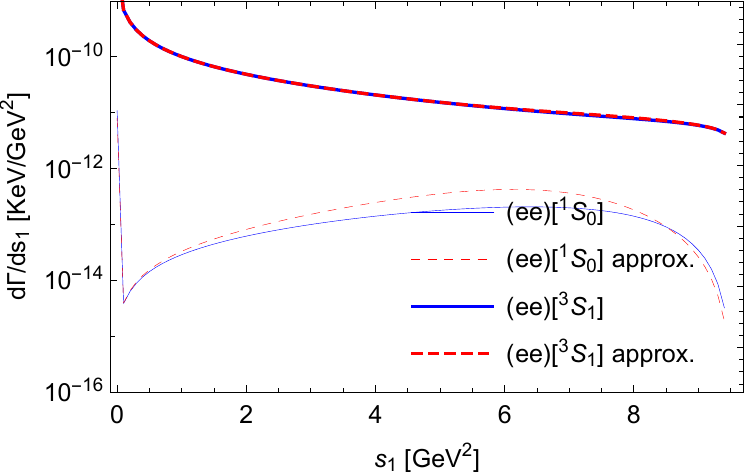}
    \includegraphics[width=0.32\linewidth]{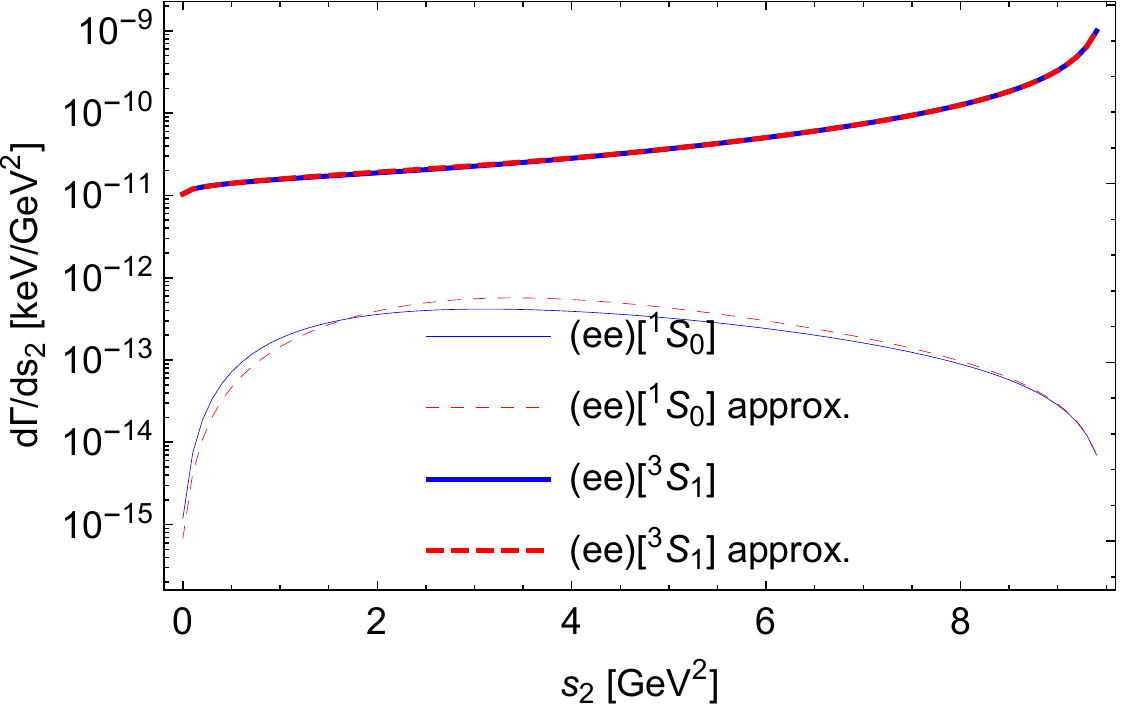}
    \includegraphics[width=0.32\linewidth]{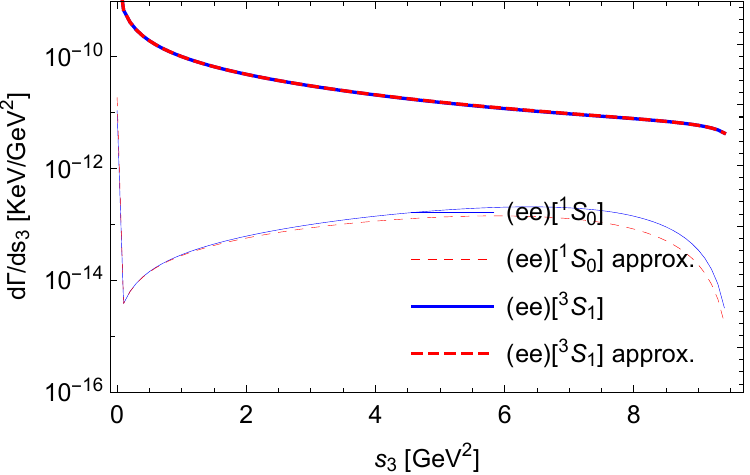}
    \caption{Differential distributions for $J/\psi \longrightarrow (e^+ e^-)[n] + e^++ e^-$ versus the invariants $s_1$, $s_2$, and $s_3$. Curves labeled ``approx.'' correspond to the $m_e^2 \ll m_{J/\psi}^2$ limit.}
    \label{fig:dGds123ee}
\end{figure}

In Figs. \ref{fig:dGds123uu} and \ref{fig:dGds123ee}, we present the differential distributions $d\Gamma/ds_{1,2,3}$ for $J/\psi \longrightarrow (l^+ l^-)[n] + l^+ + l^-$ ($l=\mu,\,e$), where the approximated curves correspond to Eqs. \eqref{eq:sqM1S0l} and \eqref{eq:sqM3S1l} in the $m_l^2 \ll m_{J/\psi}^2$ limit. Distributions $d\Gamma/ds_1$ and $d\Gamma/ds_3$ are identical because $l^{+}(p_2)$ and $l^{-}(p_3)$ are the same lepton type with opposite charges.

Finally, we discuss the inclusive production of dimuonium $(\mu^+\mu^-)[n]$ and positronium $(e^+e^-)[n]$ in $J/\psi$ decays. Combining Eqs. \eqref{eq:jps2La-1}, \eqref{eq:jps2La-2}, \eqref{eq:jpsi2Lmue-1}–\eqref{eq:jpsi2Lemu-2}, and \eqref{eq:jpsi2Lmumu-1}–\eqref{eq:jpsi2Lee-2}, and using $\Gamma_{J/\psi} = 92.6$ keV, the inclusive branching fractions are:
\begin{align}
    Br(J/\psi \longrightarrow (\mu^+ \mu^-)[^1S_0] + X) &= 4.2 \times 10^{-13} ,\\
    Br(J/\psi \longrightarrow (\mu^+ \mu^-)[^3S_1] + X) &= 1.5 \times 10^{-12} ;\\
    Br(J/\psi \longrightarrow (e^+ e^-)[^1S_0] + X) &= 2.4 \times 10^{-14} ,\\
    Br(J/\psi \longrightarrow (e^+ e^-)[^3S_1] + X) &= 1.7 \times 10^{-11} ,
\end{align}
where $X$ denotes $\gamma$ or charged leptons $\mu^\pm,\,e^\pm$. Inclusive $(\mu^+ \mu^-)[^1S_0]$ production is dominated by $J/\psi \longrightarrow (\mu^+ \mu^-)[^1S_0] +\gamma$, while inclusive $(e^+ e^-)[^1S_0]$ production is dominated by $J/\psi \longrightarrow (e^+ e^-)[^1S_0] + e^+ e^-$. With STCF's annual production of $3.4 \times 10^{12}$ $J/\psi$ events, the observation of $(e^+ e^-)[^3S_1]$, $(\mu^+ \mu^-)[^3S_1]$, and even $(\mu^+ \mu^-)[^1S_0]$ within several years is highly promising. 

We note that dimuonium has not yet been observed experimentally. The $(\mu^+ \mu^-)[^3S_1]$ state could be discovered via its dominant decay mode $(\mu^+ \mu^-)[^3S_1]\longrightarrow e^+e^- (\gamma)$, while $(\mu^+ \mu^-)[^1S_0]$ might be discovered via its dominant $(\mu^+ \mu^-)[^1S_0]\longrightarrow 2\gamma$ decay. For $(\mu^+ \mu^-)[^3S_1]\longrightarrow e^+e^- (\gamma)$ decays, the hypothetical $X(17)$ particle \cite{Krasznahorkay:2015iga}, if it exists, could leave signatures in the invariant $e^+e^-$ spectrum.

%%%%%%%%%%%%%%%%%%%%%%%%%%%%%%%%%%%%%%%%%%%%%%%%%%%%%%%%%%%%%%%%%%%%
\section{Leptonium production in bottomonium decays}
\label{sec:bottomleptonium}
%%%%%%%%%%%%%%%%%%%%%%%%%%%%%%%%%%%%%%%%%%%%%%%%%%%%%%%%%%%%%%%%%%%%

Given that the mass of bottomonium exceeds the total mass of four tau leptons, the production of ditauonium $(\tau^+ \tau^-)[n]$, tauonium $(\tau^\pm \mu^\mp)[n]$, and $(\tau^\pm e^\mp)[n]$ is kinematically permitted. In this section, we focus on discussing the production of tau-related leptonium in $\Upsilon$ decays.

%%%%%%%%%%%%%%%%%%%%%%%%%%%%%%%%%
\subsection{$\Upsilon \longrightarrow l^+l^-$}
\label{subsec:upsilon2ll}
%%%%%%%%%%%%%%%%%%%%%%%%%%%%%%%%%

PDG2024 provides branching fractions for the decays $\Upsilon \longrightarrow l^+l^-$ ($l=\tau,\,\mu,\,e$) \cite{ParticleDataGroup:2024cfk}, from which we derive the following experimental ratios:
\begin{align}
    \left.\frac{Br(\Upsilon \longrightarrow \tau^+\tau^-)}{Br(\Upsilon \longrightarrow \mu^+\mu^-)}\right|_{\mathrm{exp}} &= 1.05 \pm 0.04, \\
    \left.\frac{Br(\Upsilon \longrightarrow \mu^+\mu^-)}{Br(\Upsilon \longrightarrow e^+e^-)}\right|_{\mathrm{exp}} &= 1.04 \pm 0.04.
\end{align}
The theoretical decay widths for $\Upsilon \longrightarrow l^+l^-$ ($l=\tau,\,\mu,\,e$) can be obtained from Eq. \eqref{eq:jpsi2ll} by simply replacing the electric charge $e_Q$, wavefunction $R(0)$, and quark mass. This yields the theoretical predictions for the ratios:
\begin{align}
    \left.\frac{Br(\Upsilon \longrightarrow \tau^+\tau^-)}{Br(\Upsilon \longrightarrow \mu^+\mu^-)}\right|_{\mathrm{the}} &= 0.99215 \pm \left(1.7 \times 10^{-6}\right), \\
    \left.\frac{Br(\Upsilon \longrightarrow \mu^+\mu^-)}{Br(\Upsilon \longrightarrow e^+e^-)}\right|_{\mathrm{the}} &= 1.00000 \pm \left(3.9 \times 10^{-12}\right),
\end{align}
where the errors arising from mass uncertainties are extremely small. The theoretical prediction for the $\tau/\mu$ ratio differs from the experimental result by 1.4$\sigma$.

Using the total decay width $\Gamma_\Upsilon = 54.02 \pm 1.25$ keV and the experimental branching fraction $Br(\Upsilon \longrightarrow \mu^+\mu^-)\left.\right|_{\mathrm{exp}} = (2.48 \pm 0.04)\%$ (the most precise among the three channels) from PDG2024 \cite{ParticleDataGroup:2024cfk}, we extract the radial wavefunction of $\Upsilon$ at the origin:
\begin{equation}
    R_{\Upsilon}^2(0) = 5.06 \pm 0.14 \,\, \mathrm{GeV}^3,
\end{equation}
which will be used in subsequent numerical evaluations.

%%%%%%%%%%%%%%%%%%%%%%%%%%%%%%%%%
\subsection{$\Upsilon \longrightarrow l^+ l^- \gamma$}
\label{subsec:upsilon2lla}
%%%%%%%%%%%%%%%%%%%%%%%%%%%%%%%%%

\begin{figure}
    \centering
    \includegraphics[width=1\linewidth]{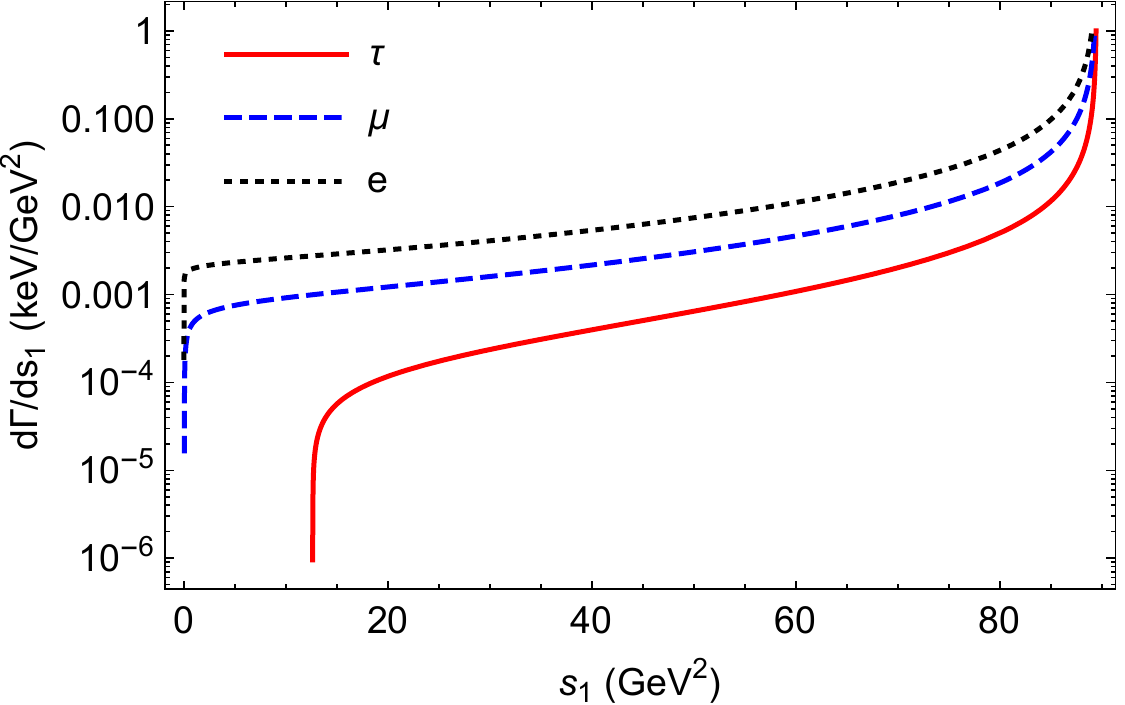}
    \caption{Differential distributions $d\Gamma\left(\Upsilon \rightarrow l^{+} l^{-} \gamma \right)/ds_1$ for taus (solid), muons (dashed)  and electrons (dotted). $s_1$ is the squared invariant mass of the lepton pair.}
    \label{fig:dGds1upsilon}
\end{figure}

The differential distributions $d\Gamma/ds_1$ for $\Upsilon \longrightarrow l^+ l^- \gamma$ ($l=\tau,\,\mu,\,e$) take the form of Eq. \eqref{eq:jpsi2lla_ds1}, with substitutions for the charge, radial wavefunction, and masses. Here, $s_1$ denotes the squared invariant mass of the lepton pair. These distributions are plotted in Fig. \ref{fig:dGds1upsilon}. The curves for taus, muons, and electrons all increase monotonically, exhibiting distinct thresholds but sharing the same endpoint. For the $\Upsilon$ case, the approximation under the $m_l^2 \ll m_{\Upsilon}^2$ limit is so accurate that no visible differences between Eqs. \eqref{eq:jpsi2lla_ds1} and \eqref{eq:jpsi2lla_ds1approxm} can be discerned in the curves, so the approximated results are not explicitly shown.

Using the total decay width $\Gamma_\Upsilon = 54.02 \pm 1.25$ keV, the branching fractions for $\Upsilon$ decays into a lepton pair and a photon with the cut $E_\gamma > 100$ MeV are:
\begin{align}
        Br\left(\Upsilon \rightarrow \tau^{+} \tau^{-} \gamma \right) &= 0.00069 \pm 0.00036, \\
        Br\left(\Upsilon \rightarrow \mu^{+} \mu^{-} \gamma \right) &= 0.0028 \pm 0.0014, \\
        Br\left(\Upsilon \rightarrow e^{+} e^{-} \gamma \right) &= 0.0066 \pm 0.0033.
\end{align}
These values differ significantly due to the mass hierarchy among the three generations of leptons. Notably, no experimental measurements of these branching fractions exist to date.

%%%%%%%%%%%%%%%%%%%%%%%%%%%%%%%%%
\subsection{$\Upsilon \longrightarrow (l^+ l^-)[^1S_0] + \gamma$}
\label{subsec:Upsilon2leptoniuma}
%%%%%%%%%%%%%%%%%%%%%%%%%%%%%%%%%

The decay width for $\Upsilon$ decays into leptonium and a photon can be obtained from Eq. \eqref{eq:jpsi2La} with appropriate substitutions. The branching fraction for $\Upsilon \longrightarrow (\tau^+ \tau^-)[^1S_0] + \gamma$ is:
\begin{equation}
    Br(\Upsilon \longrightarrow (\tau^+ \tau^-)[^1S_0] + \gamma) = \left(4.3 \pm 1.4\right)\times 10^{-12}. \label{eq:Upsilon2ditaua}
\end{equation}
Given the $1.02\times 10^8$ $\Upsilon(1S)$ events accumulated at Belle \cite{Belle:2012iwr}, observing ditauonium via the $\Upsilon \longrightarrow (\tau^+ \tau^-)[^1S_0] + \gamma$ channel is not feasible.

The ratios of branching fractions for $\Upsilon$ decays into different dileptonia in association with a photon are:
\begin{align}
    \frac{Br(\Upsilon \longrightarrow (\tau^+ \tau^-)[^1S_0] + \gamma)}{Br(\Upsilon \longrightarrow (\mu^+ \mu^-)[^1S_0] + \gamma)} &= \frac{m_\tau^2 (m_{\Upsilon}^2 -4 m_\tau^2)}{m_\mu^2 (m_{\Upsilon}^2 -4 m_\mu^2)} = \left(0.859 \pm 0.003 \right)\frac{m^2_\tau}{m^2_\mu}, \label{eq:Upsilon2dimua} \\
    \frac{Br(\Upsilon \longrightarrow (\mu^+ \mu^-)[^1S_0] + \gamma)}{Br(\Upsilon \longrightarrow (e^+ e^-)[^1S_0] + \gamma)} &= \frac{m_\mu^2 (m_{\Upsilon}^2 -4 m_\mu^2)}{m_e^2 (m_{\Upsilon}^2 -4 m_e^2)} = \left(0.99950 \pm 0.00001 \right)\frac{m^2_\mu}{m^2_e}. \label{eq:Upsilon2diea}
\end{align}
From these, the ratio of the ditauonium branching fraction to the dimuonium branching fraction is approximately 243, while the ratio of the dimuonium branching fraction to the positronium branching fraction is approximately 42732.

%%%%%%%%%%%%%%%%%%%%%%%%%%%%%%%%%
\subsection{$\Upsilon \longrightarrow (l_1^+ l_2^-)[n] + l_1^-l_2^+$}
\label{subsec:Upsilon2leptoniumall}
%%%%%%%%%%%%%%%%%%%%%%%%%%%%%%%%%

\begin{figure}
    \centering
    \includegraphics[width=0.32\linewidth]{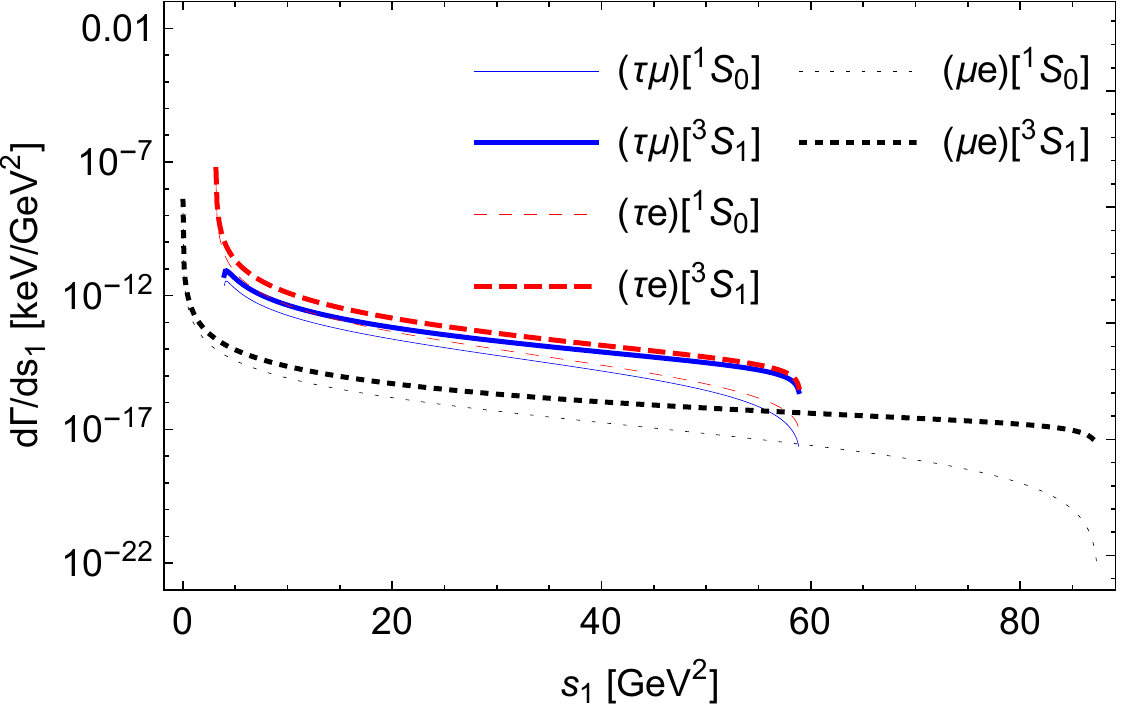}
    \includegraphics[width=0.32\linewidth]{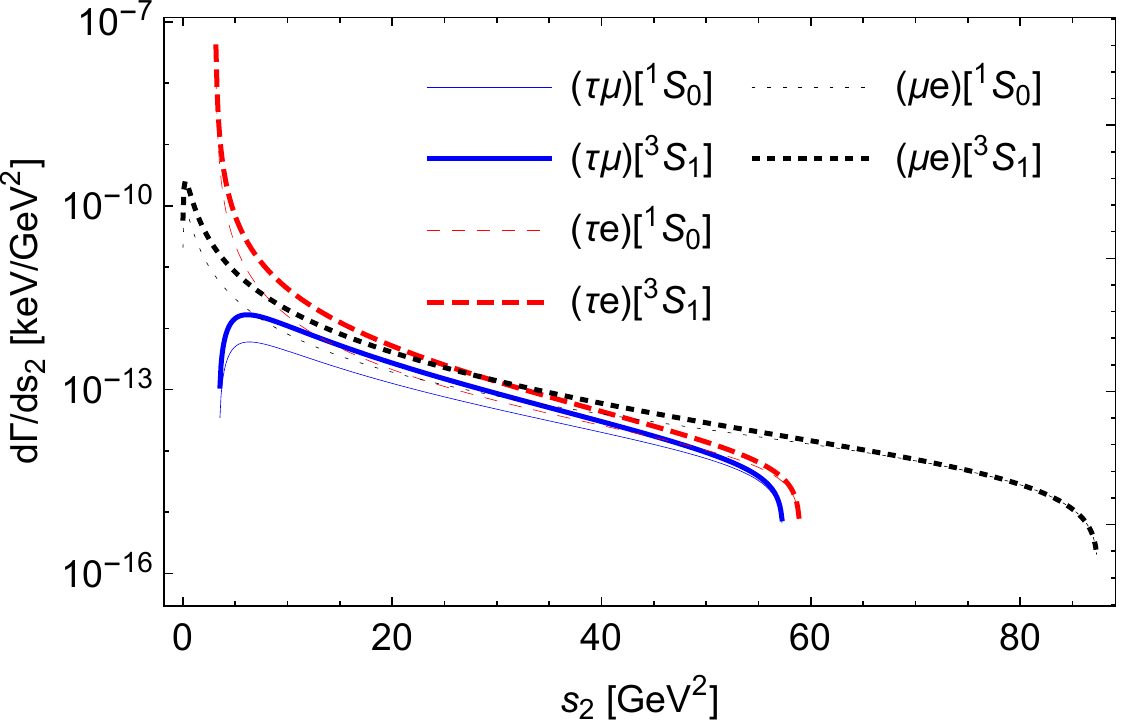}
    \includegraphics[width=0.32\linewidth]{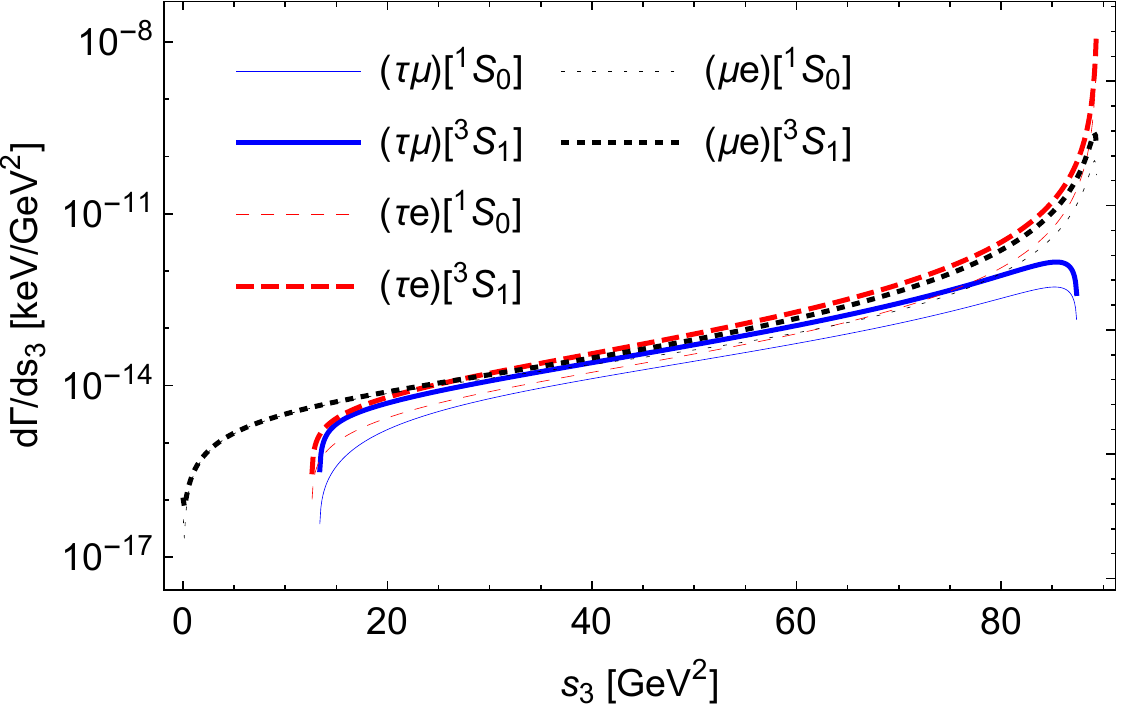}
    \caption{Differential distributions $d\Gamma/ds_{1,2,3}$ for $\Upsilon \longrightarrow (l_1^+ l_2^-)[n] + l_1^-l_2^+$ with $l_1 \neq l_2$. Thin lines and thick lines denote $^1S_0$ and $^3S_1$ states, respectively.}
    \label{fig:upsilon2l1l2d}
\end{figure}

We first discuss the production of $(\tau^+\mu^-)[n]$, $(\tau^+e^-)[n]$, and $(\mu^+e^-)[n]$ in $\Upsilon$ decays. Using the total decay width $\Gamma_\Upsilon = 54.02 \pm 1.25$ keV, their branching fractions are:
\begin{align}
    Br(\Upsilon \longrightarrow (\tau^+ \mu^-)[^1S_0] + \mu^+\tau^-) &= 1.27 \times 10^{-13},\\
    Br(\Upsilon \longrightarrow (\tau^+ \mu^-)[^3S_1] + \mu^+\tau^-) &= 3.10 \times 10^{-13}; \\
    Br(\Upsilon \longrightarrow (\tau^+ e^-)[^1S_0] + e^+\tau^-) &= 3.63 \times 10^{-11},\\
    Br(\Upsilon \longrightarrow (\tau^+ e^-)[^3S_1] + e^+\tau^-) &= 1.08 \times 10^{-10}; \\
    Br(\Upsilon \longrightarrow (\mu^+ e^-)[^1S_0] + e^+\mu^-) &= 2.20 \times 10^{-12},\\
    Br(\Upsilon \longrightarrow (\mu^+ e^-)[^3S_1] + e^+\mu^-) &= 6.27 \times 10^{-12}.
\end{align}
The branching fractions of $^3S_1$ states are roughly three times those of $^1S_0$ states, consistent with the $J/\psi$ case. Given that the $1.02\times 10^8$ $\Upsilon$ events at Belle \cite{Belle:2012iwr} are insufficient for experimental measurements, we will not further discuss event counts hereafter.
In Fig. \ref{fig:upsilon2l1l2d}, we present the differential distributions $d\Gamma/ds_{1,2,3}$ for $\Upsilon \longrightarrow (l_1^+ l_2^-)[n](p_1) + l_1^-(p_2) + l_2^+(p_3)$ with $l_1 \neq l_2$, where $s_1=(p_1+p_2)^2$, $s_2=(p_2+p_3)^2$, and $s_3=(p_1+p_3)^2$. Thin lines and thick lines correspond to $^1S_0$ and $^3S_1$ states, respectively. As expected, the line shapes are similar to those for the $J/\psi$ case in Fig. \ref{fig:dGds123ue}.

\begin{figure}
    \centering
    \includegraphics[width=0.32\linewidth]{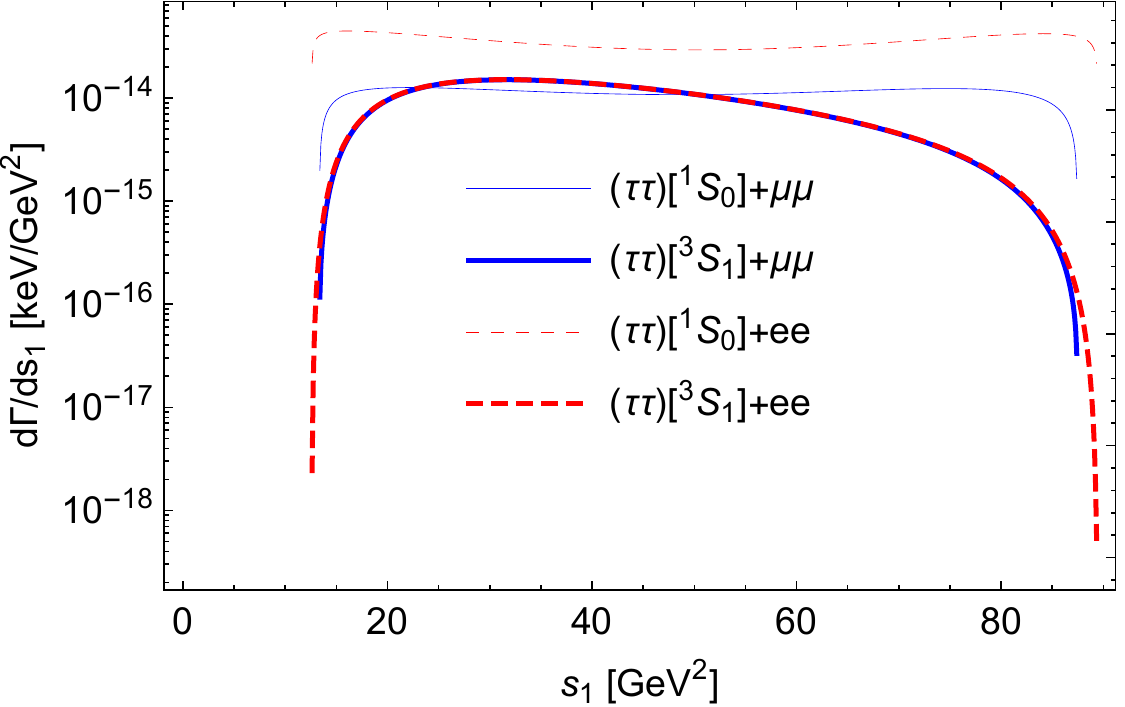}
    \includegraphics[width=0.32\linewidth]{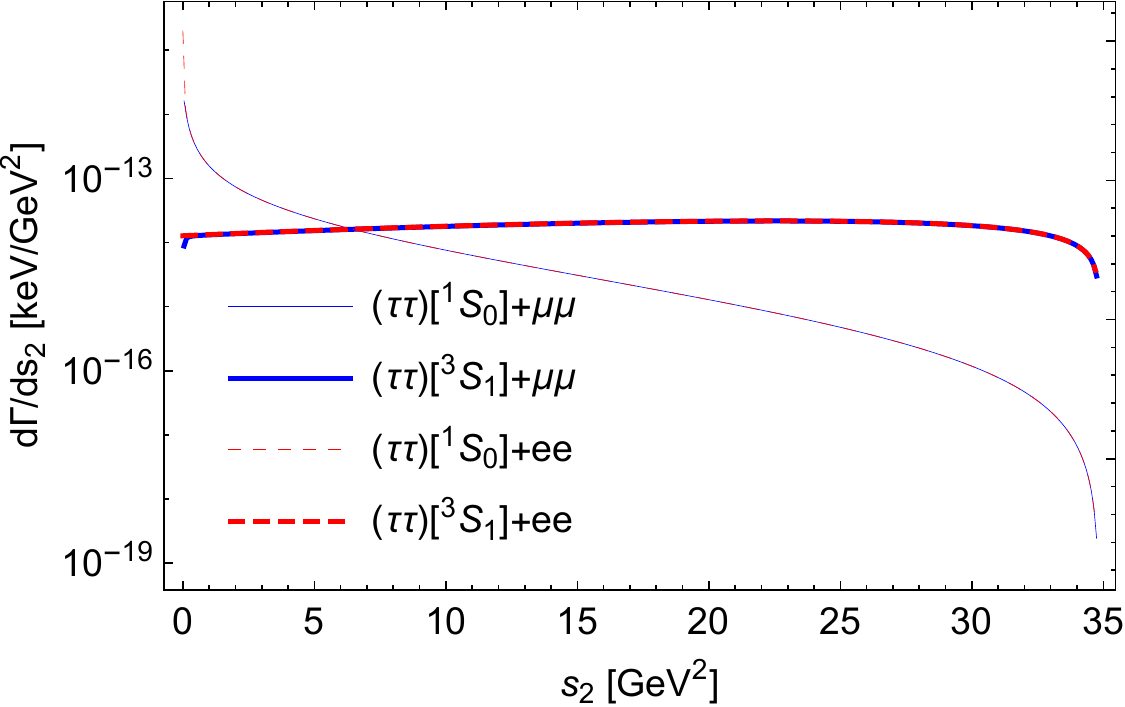}
    \includegraphics[width=0.32\linewidth]{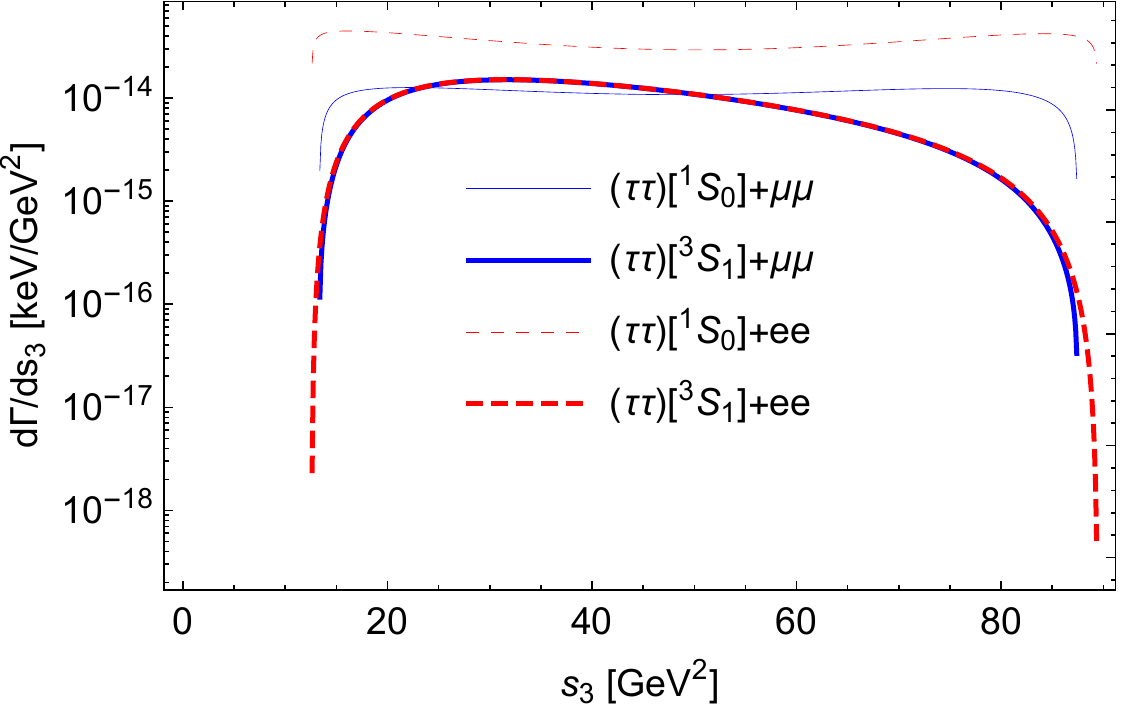}
    \caption{Differential distributions $d\Gamma/ds_{1,2,3}$ for $\Upsilon \longrightarrow (\tau^+\tau^-)[n] + l^+l^-$ ($l=\mu,e$).}
    \label{fig:upsilon2ditaud}
\end{figure}

For the production of $\Upsilon \longrightarrow (\tau^+\tau^-)[n] + \mu^+ \mu^-$ and $\Upsilon \longrightarrow (\tau^+\tau^-)[n] + e^+e^-$, the branching fractions are:
\begin{align}
    Br(\Upsilon \longrightarrow (\tau^+\tau^-)[^1S_0] + \mu^+ \mu^-) &= 1.55 \times 10^{-14}, \label{eq:Upsilon2ditaumu1S0} \\
    Br(\Upsilon \longrightarrow (\tau^+\tau^-)[^3S_1] + \mu^+ \mu^-) &= 1.16 \times 10^{-14}; \label{eq:Upsilon2ditaumu3S1} \\
    Br(\Upsilon \longrightarrow (\tau^+\tau^-)[^1S_0] + e^+ e^-) &= 5.08 \times 10^{-14}, \label{eq:Upsilon2ditaue1S0} \\
    Br(\Upsilon \longrightarrow (\tau^+\tau^-)[^3S_1] + e^+ e^-) &= 1.16 \times 10^{-14}.\label{eq:Upsilon2ditaue3S1}
\end{align}
Notably, the branching fractions of $^3S_1$ states are smaller than those of $^1S_0$ states. This arises because $^1S_0$ and $^3S_1$ states originate from distinct Feynman diagram topologies, specifically the 3rd and 4th diagrams in Fig. \ref{fig:jpsi2leptoniumll}, respectively. This behavior contrasts with $J/\psi \longrightarrow (\mu^+ \mu^-)[n] + e^+e^-$ and $J/\psi \longrightarrow (e^+ e^-)[n] + \mu^+\mu^-$, where $^3S_1$ branching fractions exceed those of $^1S_0$.
In Fig. \ref{fig:upsilon2ditaud}, we show the differential distributions $d\Gamma/ds_{1,2,3}$ for $\Upsilon \longrightarrow (\tau^+\tau^-)[n] + l^+l^-$ ($l=\mu,e$). States with the same spin exhibit consistent trends, and $d\Gamma/ds_1$ distributions match $d\Gamma/ds_3$ because the outgoing leptons are same-generation particles with opposite charges.

\begin{figure}
    \centering
    \includegraphics[width=0.32\linewidth]{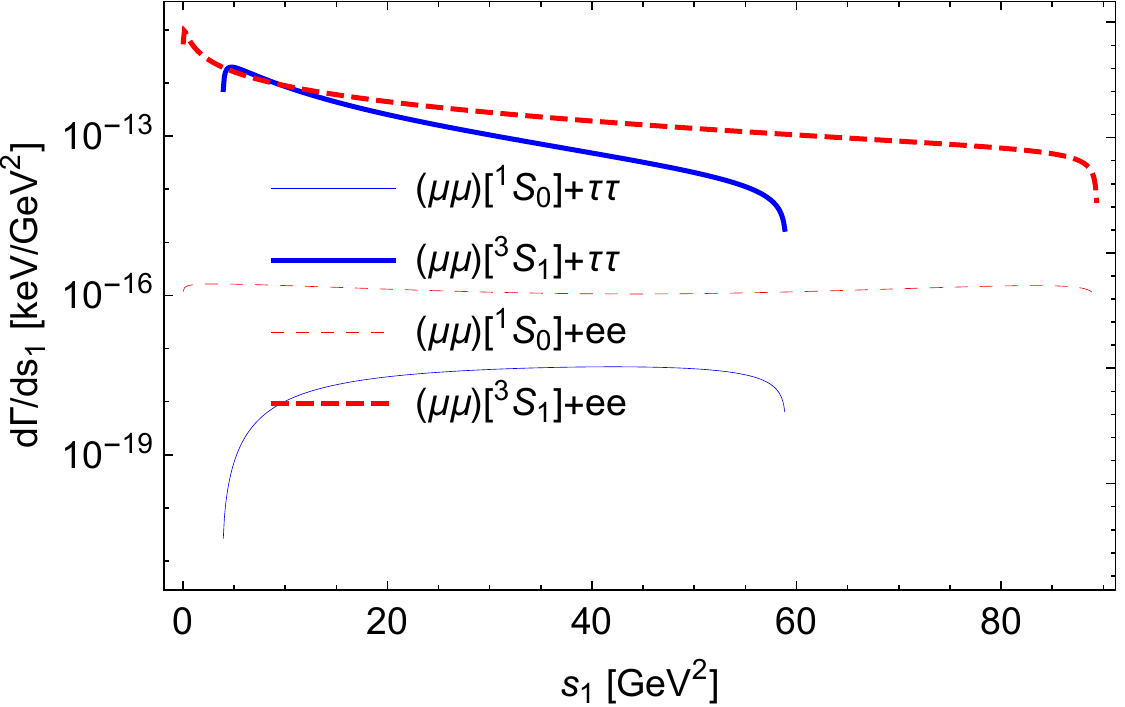}
    \includegraphics[width=0.32\linewidth]{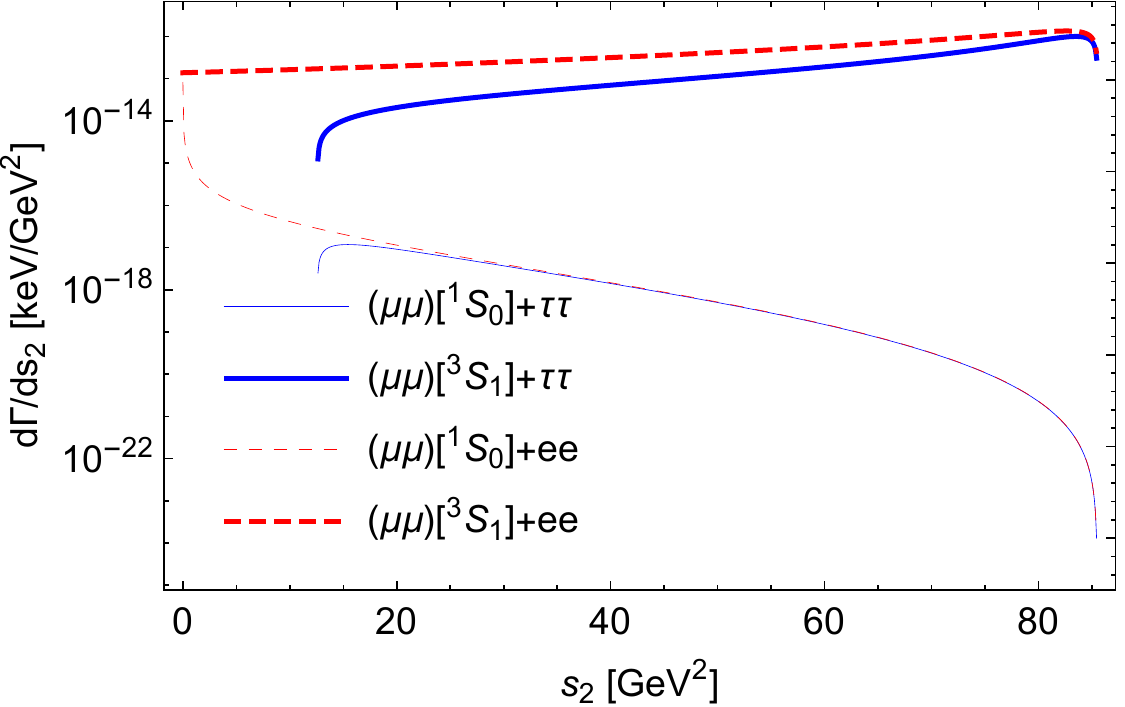}
    \includegraphics[width=0.32\linewidth]{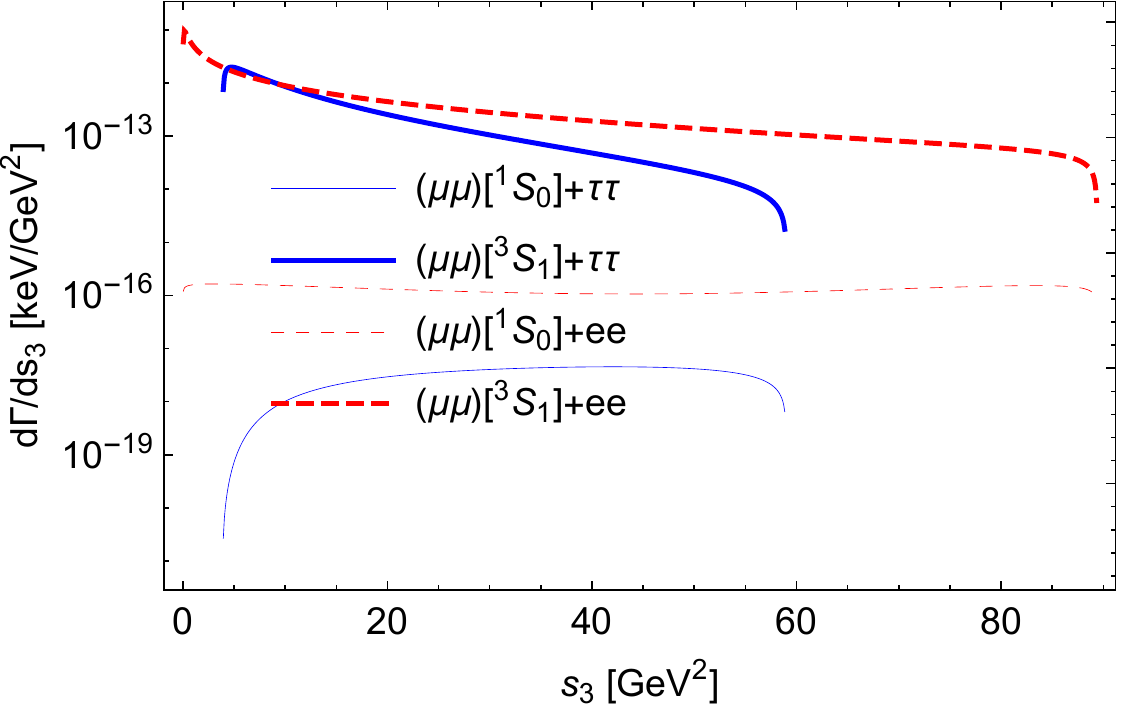}
    \caption{Differential distributions $d\Gamma/ds_{1,2,3}$ for $\Upsilon \longrightarrow (\mu^+\mu^-)[n] + l^+l^-$ ($l=\tau,e$).}
    \label{fig:upsilon2dimud}
\end{figure}

For the production of $\Upsilon \longrightarrow (\mu^+ \mu^-)[n] + \tau^+\tau^-$ and $\Upsilon \longrightarrow (\mu^+ \mu^-)[n] + e^+e^-$, the branching fractions are:
\begin{align}
    Br(\Upsilon \longrightarrow (\mu^+\mu^-)[^1S_0] + \tau^+\tau^-) &= 3.21 \times 10^{-18}, \label{eq:Upsilon2dimutau1S0} \\
    Br(\Upsilon \longrightarrow (\mu^+\mu^-)[^3S_1] + \tau^+\tau^-) &= 2.99 \times 10^{-13}; \label{eq:Upsilon2dimutau3S1}\\
    Br(\Upsilon \longrightarrow (\mu^+\mu^-)[^1S_0] + e^+ e^-) &= 2.14 \times 10^{-16}, \label{eq:Upsilon2dimue1S0}\\
    Br(\Upsilon \longrightarrow (\mu^+\mu^-)[^3S_1] + e^+ e^-) &= 7.52 \times 10^{-13}. \label{eq:Upsilon2dimue3S1}
\end{align}
Here, $^3S_1$ branching fractions are orders of magnitude larger than those of $^1S_0$ states, indicating that the 4th diagram in Fig. \ref{fig:jpsi2leptoniumll} dominates over the 3rd diagram for $\Upsilon \longrightarrow (\mu^+\mu^-)[n] + l^+l^-$ ($l=\tau,e$).
In Fig. \ref{fig:upsilon2dimud}, we present the differential distributions $d\Gamma/ds_{1,2,3}$ for these processes, with $d\Gamma/ds_1$ matching $d\Gamma/ds_3$ as expected.

\begin{figure}
    \centering
    \includegraphics[width=0.32\linewidth]{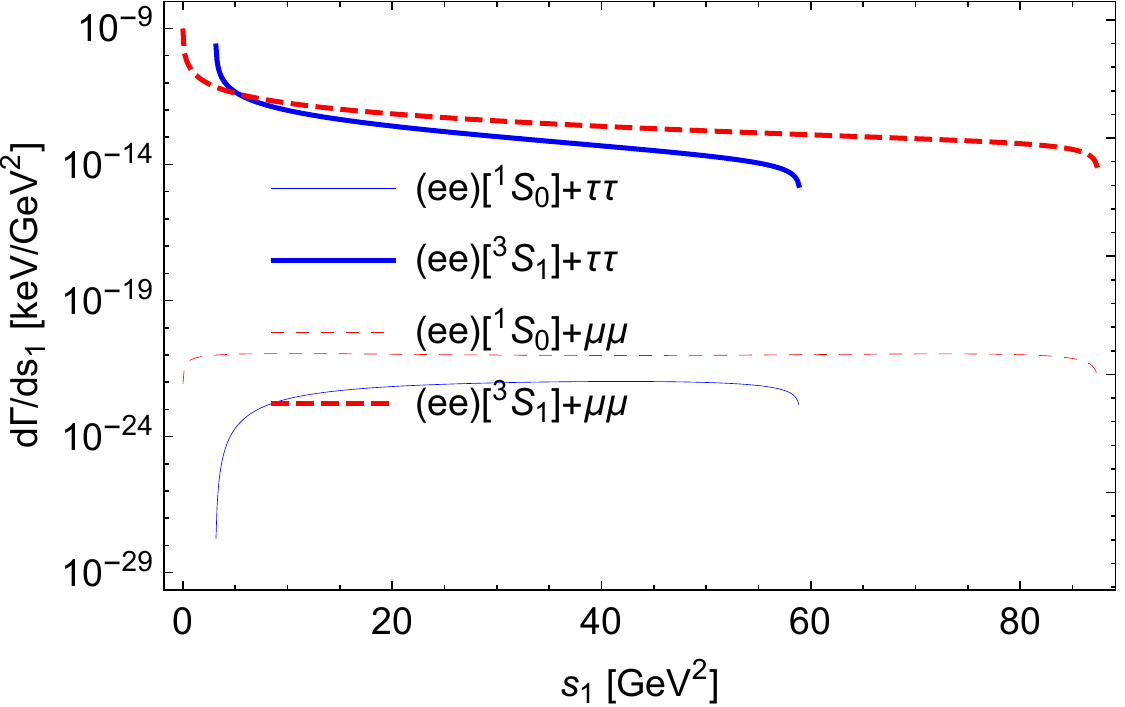}
    \includegraphics[width=0.32\linewidth]{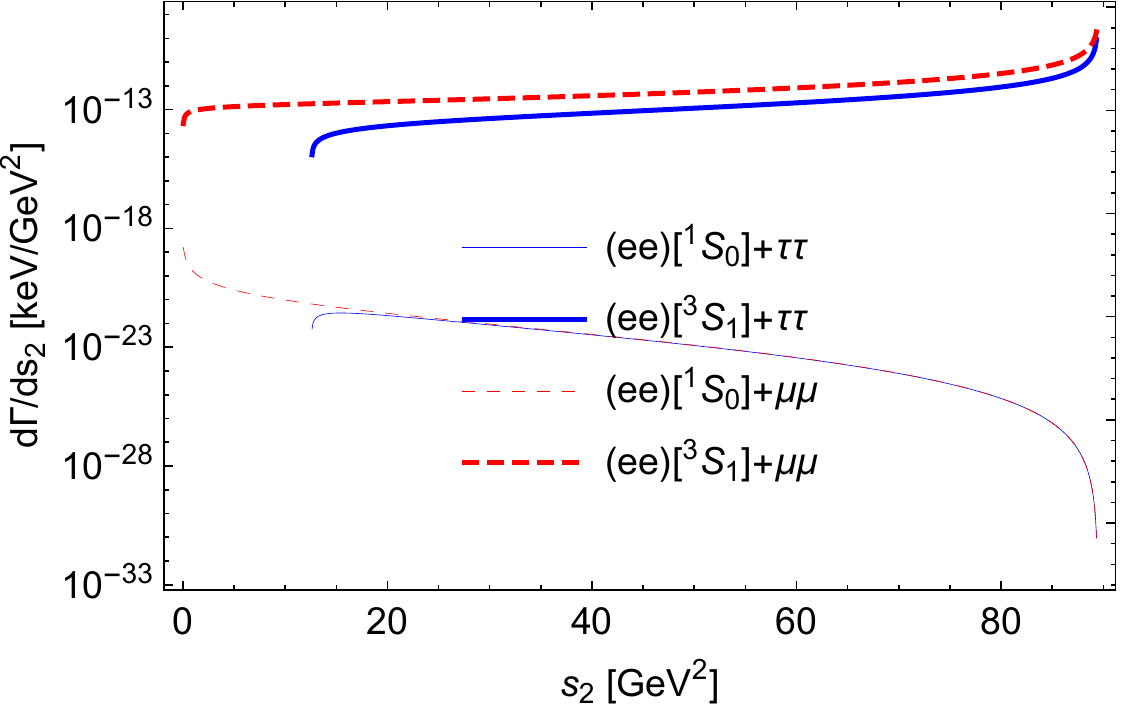}
    \includegraphics[width=0.32\linewidth]{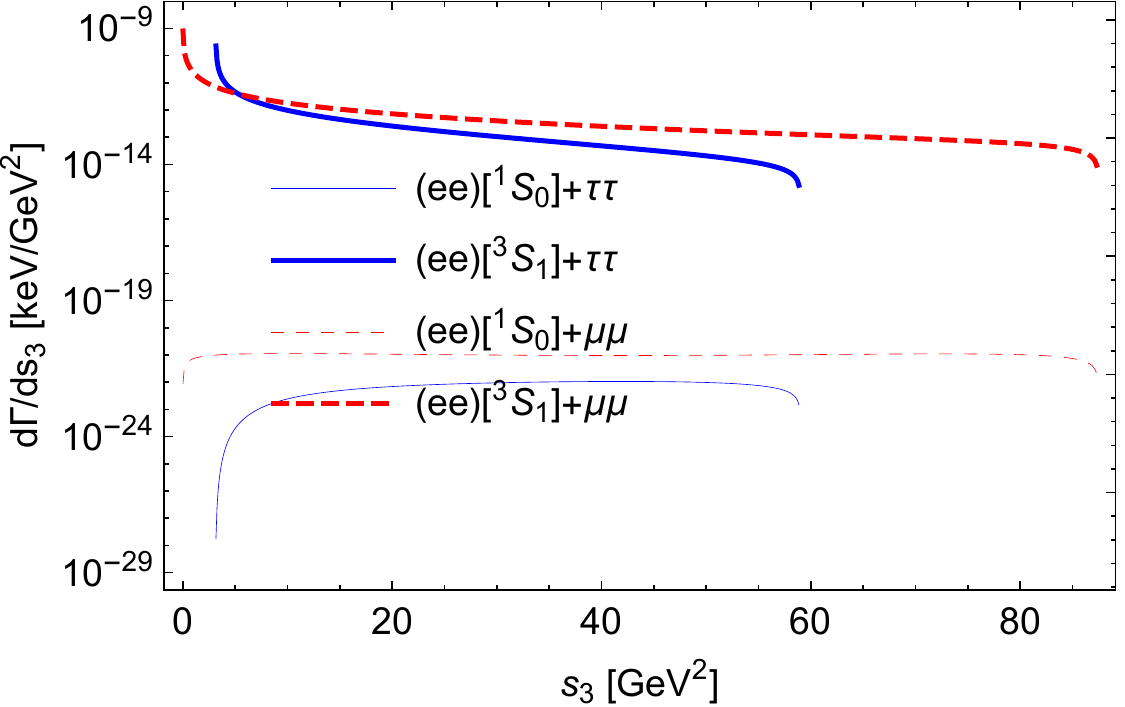}
    \caption{Differential distributions $d\Gamma/ds_{1,2,3}$ for $\Upsilon \longrightarrow (e^+ e^-)[n] + l^+l^-$ ($l=\tau,\mu$).}
    \label{fig:upsilon2died}
\end{figure}

For the production of $\Upsilon \longrightarrow (e^+ e^-)[n] + \tau^+\tau^-$ and $\Upsilon \longrightarrow (e^+ e^-)[n] + \mu^+\mu^-$, the branching fractions are:
\begin{align}
    Br(\Upsilon \longrightarrow (e^+e^-)[^1S_0] + \tau^+\tau^-) &= 7.54 \times 10^{-23}, \label{eq:Upsilon2dietau1S0}\\
    Br(\Upsilon \longrightarrow (e^+e^-)[^3S_1] + \tau^+\tau^-) &= 1.26 \times 10^{-12}; \label{eq:Upsilon2dietau3S1}\\
    Br(\Upsilon \longrightarrow (e^+e^-)[^1S_0] + \mu^+ \mu^-) &= 1.62 \times 10^{-21}, \label{eq:Upsilon2diemu1S0}\\
    Br(\Upsilon \longrightarrow (e^+e^-)[^3S_1] + \mu^+ \mu^-) &= 3.98 \times 10^{-12}. \label{eq:Upsilon2diemu3S1}
\end{align}
Similar to $\Upsilon \longrightarrow (\mu^+\mu^-)[n] + l^+l^-$ ($l=\tau,e$), $^3S_1$ branching fractions are orders of magnitude larger than those of $^1S_0$ states.
In Fig. \ref{fig:upsilon2died}, we show the differential distributions $d\Gamma/ds_{1,2,3}$ for $\Upsilon \longrightarrow (e^+ e^-)[n] + l^+l^-$ ($l=\tau,\mu$), with $d\Gamma/ds_1$ matching $d\Gamma/ds_3$.

\begin{figure}
    \centering
    \includegraphics[width=0.32\linewidth]{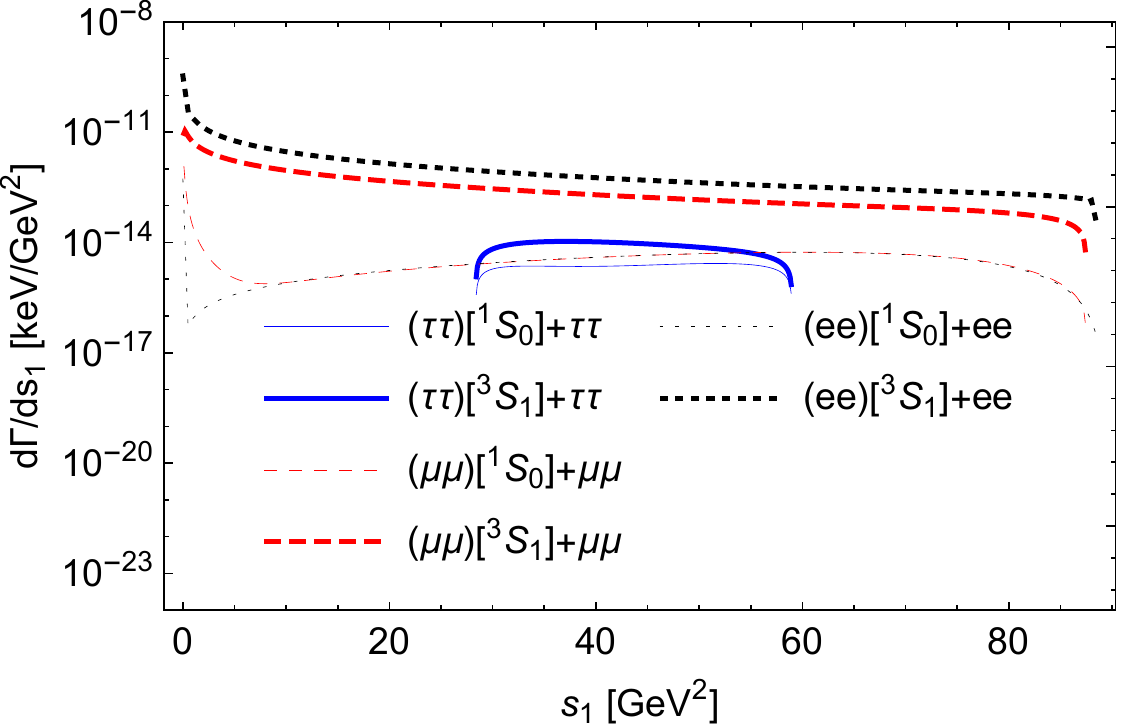}
    \includegraphics[width=0.32\linewidth]{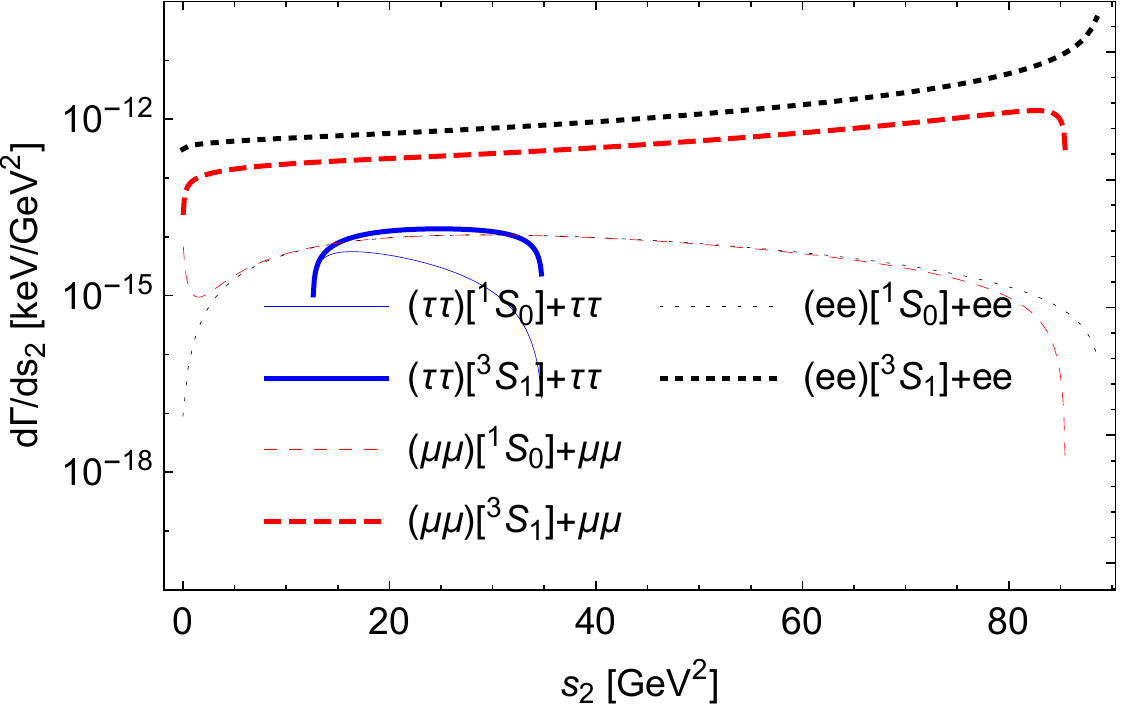}
    \includegraphics[width=0.32\linewidth]{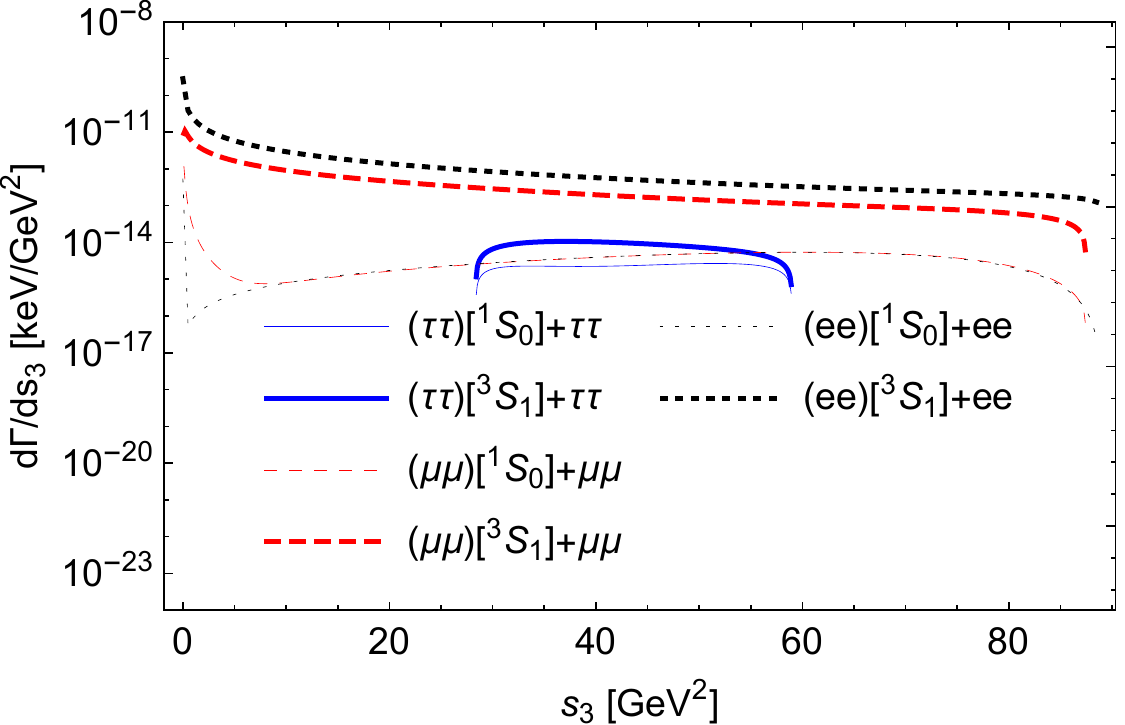}
    \caption{Differential distributions $d\Gamma/ds_{1,2,3}$ for $\Upsilon \longrightarrow (l^+l^-)[n] + l^+ l^-$ ($l=\tau,\mu,e$).}
    \label{fig:upsilon4ld}
\end{figure}

For the production of $\Upsilon \longrightarrow (l^+l^-)[n] + l^+ l^-$ ($l=\tau,\mu,e$), the branching fractions are:
\begin{align}
    Br(\Upsilon \longrightarrow (\tau^+ \tau^-)[^1S_0] + \tau^+\tau^-) &= 1.28 \times 10^{-15}, \label{eq:Upsilon24tau1S0}\\
    Br(\Upsilon \longrightarrow (\tau^+ \tau^-)[^3S_1] + \tau^+\tau^-) &= 4.49 \times 10^{-15}; \label{eq:Upsilon24tau3S1}\\
    Br(\Upsilon \longrightarrow (\mu^+ \mu^-)[^1S_0] + \mu^+\mu^-) &= 9.68 \times 10^{-15}, \label{eq:Upsilon24mu1S0}\\
    Br(\Upsilon \longrightarrow (\mu^+ \mu^-)[^3S_1] + \mu^+\mu^-) &= 7.77 \times 10^{-13}; \label{eq:Upsilon24mu3S1}\\
    Br(\Upsilon \longrightarrow (e^+ e^-)[^1S_0] + e^+e^-) &= 9.83 \times 10^{-15}, \label{eq:Upsilon24e1S0}\\
    Br(\Upsilon \longrightarrow (e^+ e^-)[^3S_1] + e^+e^-) &= 5.76 \times 10^{-12}. \label{eq:Upsilon24e3S1}
\end{align}
All four diagrams in Fig. \ref{fig:jpsi2leptoniumll} contribute here, with the 3rd diagram vanishing for $^3S_1$ states and the 4th diagram vanishing for $^1S_0$ states. Branching fractions for $^3S_1$ states are significantly larger than those for $^1S_0$ states.
In Fig. \ref{fig:upsilon4ld}, we present the differential distributions $d\Gamma/ds_{1,2,3}$ for these processes, with $d\Gamma/ds_1$ matching $d\Gamma/ds_3$.

Finally, we discuss the inclusive production of ditauonium $(\tau^+ \tau^-)[n]$, dimuonium $(\mu^+\mu^-)[n]$, and positronium $(e^+e^-)[n]$ in $\Upsilon$ decays. Combining results from Eqs. \eqref{eq:Upsilon2ditaua}–\eqref{eq:Upsilon2diea} and \eqref{eq:Upsilon2ditaumu1S0}–\eqref{eq:Upsilon24e3S1}, the inclusive branching fractions are:
\begin{align}
    Br(\Upsilon \longrightarrow (\tau^+ \tau^-)[^1S_0] + X) &= 4.3 \times 10^{-12} ,\\
    Br(\Upsilon \longrightarrow (\tau^+ \tau^-)[^3S_1] + X) &= 2.8 \times 10^{-14} ;\\
    Br(\Upsilon \longrightarrow (\mu^+ \mu^-)[^1S_0] + X) &= 2.7 \times 10^{-14} ,\\
    Br(\Upsilon \longrightarrow (\mu^+ \mu^-)[^3S_1] + X) &= 1.8 \times 10^{-12} ;\\
    Br(\Upsilon \longrightarrow (e^+ e^-)[^1S_0] + X) &= 9.9 \times 10^{-15} ,\\
    Br(\Upsilon \longrightarrow (e^+ e^-)[^3S_1] + X) &= 1.1 \times 10^{-11} ,
\end{align}
where $X$ denotes $\gamma$ or charged leptons. Inclusive $(\tau^+ \tau^-)[^1S_0]$ production is dominated by the radiative process $\Upsilon \longrightarrow (\tau^+ \tau^-)[^1S_0] + \gamma$. The inclusive branching fractions of $(\mu^+ \mu^-)[^3S_1] + X$ and $(e^+ e^-)[^3S_1] + X$ in $\Upsilon$ decays are of the same order of magnitude as those in $J/\psi$ decays.

%%%%%%%%%%%%%%%%%%%%%%%%%%%%%%%%%%%%%%%%%%
\section{SUMMARY}
\label{sec:summary}
%%%%%%%%%%%%%%%%%%%%%%%%%%%%%%%%%%%%%%%%%%

In this paper, we investigate leptonium production in heavy quarkonium decays within the non-relativistic QED/QCD framework. We discuss the processes \({\cal{Q}} \longrightarrow (l_1^+ l_2^-)[n] +\gamma\) (where \(l_{1,2}=\tau,\,\mu,\,e\)) and \({\cal{Q}} \longrightarrow (l_1^+ l_2^-)[n] + l_1^- l_2^+\), where \({\cal{Q}}\) denotes the heavy quarkonia \(J/\psi\) or \(\Upsilon\), and \(n= {^1S_0}\) or \(^3S_1\) correspond to para-leptonium and ortho-leptonium, respectively. We analyze the branching fractions and invariant-mass distributions of the three-body decays.

Given the annual production of \(3.4 \times 10^{12}\) \(J/\psi\) events at the future Super Tau-Charm Facility (STCF), prospects for experimental observations of leptonium at STCF are promising. Processes with branching fractions of order \(\sim 10^{-12}\) include \(J/\psi \longrightarrow (\mu^+\mu^-)[^3S_1] +X\) and \(J/\psi \longrightarrow (\mu^+e^-)[^1S_0] + e^+ \mu^-\), while those with branching fractions of order \(\sim 10^{-11}\) include \(J/\psi \longrightarrow (e^+e^-)[^3S_1] +X\) and \(J/\psi \longrightarrow (\mu^+e^-)[^3S_1] + e^+ \mu^-\). Since dimuonium has not yet been observed experimentally, the inclusive process \(J/\psi \longrightarrow (\mu^+\mu^-)[^3S_1] +X\) offers a valuable opportunity. The ortho-dimuonium could be discovered via its dominant decay mode \((\mu^+ \mu^-)[^3S_1]\longrightarrow e^+e^- (\gamma)\) at future STCF, and this channel may also be used to probe the existence of the hypothetical \(X(17)\) particle in the $e^+e^-$ spectrum.

The production of tauonium is kinematically allowed in \(\Upsilon\) decays. Inclusive production of para-ditauonium \((\tau^+\tau^-)[^1S_0]\) is dominated by the process \(\Upsilon \longrightarrow (\tau^+ \tau^-)[^1S_0] + \gamma\) with a branching fraction of \(4.3 \times 10^{-12}\). However, insufficient \(\Upsilon\) events have been accumulated to enable experimental measurements thus far. Additionally, the two largest branching fractions for tauonium production are those of \(\Upsilon \longrightarrow (\tau^+ e^-)[^1S_0] + e^+\tau^-\) and \(\Upsilon \longrightarrow (\tau^+ e^-)[^3S_1] + e^+\tau^-\), which are of order \(\sim 10^{-11}\) and \(\sim 10^{-10}\), respectively. The decay modes of tauonium \((\tau^+ e^-)[n]\) are dominated by weak decays of its constituent \(\tau^+\) lepton.

In this work, we investigate the production of leptonium in the rare decays of heavy quarkonia, specifically $J/\psi$ and $\Upsilon$. This production process is highly dependent on the availability and size of heavy quarkonium datasets, which represents a key limitation of experimental studies.  
Notably, this limitation can be significantly mitigated by leveraging the large $J/\psi$ dataset that will be available at the future STCF. Additionally, uncertainties arise in other production mechanisms: photoproduction, electroproduction, and ultraperipheral $AA$ collisions at hadron colliders, as well as two-photon fusion at $e^+e^-$ colliders, all require the use of nonperturbative photon luminosity functions—for protons/nuclei ($p/A$) and electrons, respectively—introducing unavoidable uncertainties.  
In contrast, heavy quarkonium decays offer a distinct advantage: the nonperturbative wavefunction of heavy quarkonia can be eliminated by measuring the ratios of decay rates between two different channels. With the forthcoming large $J/\psi$ dataset at STCF and the $\Upsilon$ dataset at Belle II, we thus argue that a unique and promising opportunity will emerge to study leptonium in heavy quarkonium decays.  

%%%%%%%%%%%%%%%%%%%%%%%%%%%%%%%%%%%%%%%%%%%%%%%%%%%%%%%%%%%%%%%%%%%%%
\vspace{0.5cm} {\bf Acknowledgments}
This work is supported in part by the National Natural Science Foundation of China (NSFC) under the Grants No. 12475083.
%%%%%%%%%%%%%%%%%%%%%%%%%%%%%%%%%%%%%%%%%%%%%%%%%%%%%%%%%%%%%%%%%%%%

\bibliography{Jpsi2leptonium}

\end{document}